\immediate\write18{makeindex \jobname.nlo -s nomencl.ist -o \jobname.nls} 
\documentclass[dissertation, amsmath, amssymb]{aaltoseries}
\usepackage[utf8]{inputenc}
\usepackage{amssymb, amsmath, mathtools}
\usepackage[tight,footnotesize]{subfigure}
\usepackage{enumitem}
\setlist[itemize]{noitemsep, topsep=0pt}
\usepackage{pdfpages}
\usepackage[english]{babel}
\makeatletter

\newcommand{\Rmnum}[1]{\expandafter\@slowromancap\romannumeral #1@}
\makeatother
\usepackage{nomencl}
\renewcommand{\nompreamble}{This section shows frequently used symbols in this dissertation, especially for developing numerical simulation models and interpreting their results in Chapters~\ref{sec:Models} and~\ref{sec:Results}.} 
\makenomenclature
\usepackage{etoolbox}
\renewcommand\nomgroup[1]{%
	\item[\bfseries
	\ifstrequal{#1}{A}{Social Force Model}
		{%
		\ifstrequal{#1}{B}{Joining Behavior Model}
			{%
			\ifstrequal{#1}{C}{Modeling Pedestrian Flow near an Attraction}
				{%
				\ifstrequal{#1}{D}{Collective Behavior Measures}
					{%
					\ifstrequal{#1}{E}{Other Symbols}{}
					}
				}
			}
		}
	]
	\vspace{0.4in} 
}

\begin{document}
\includepdf[pages=1-4, noautoscale=true, width=\paperwidth]{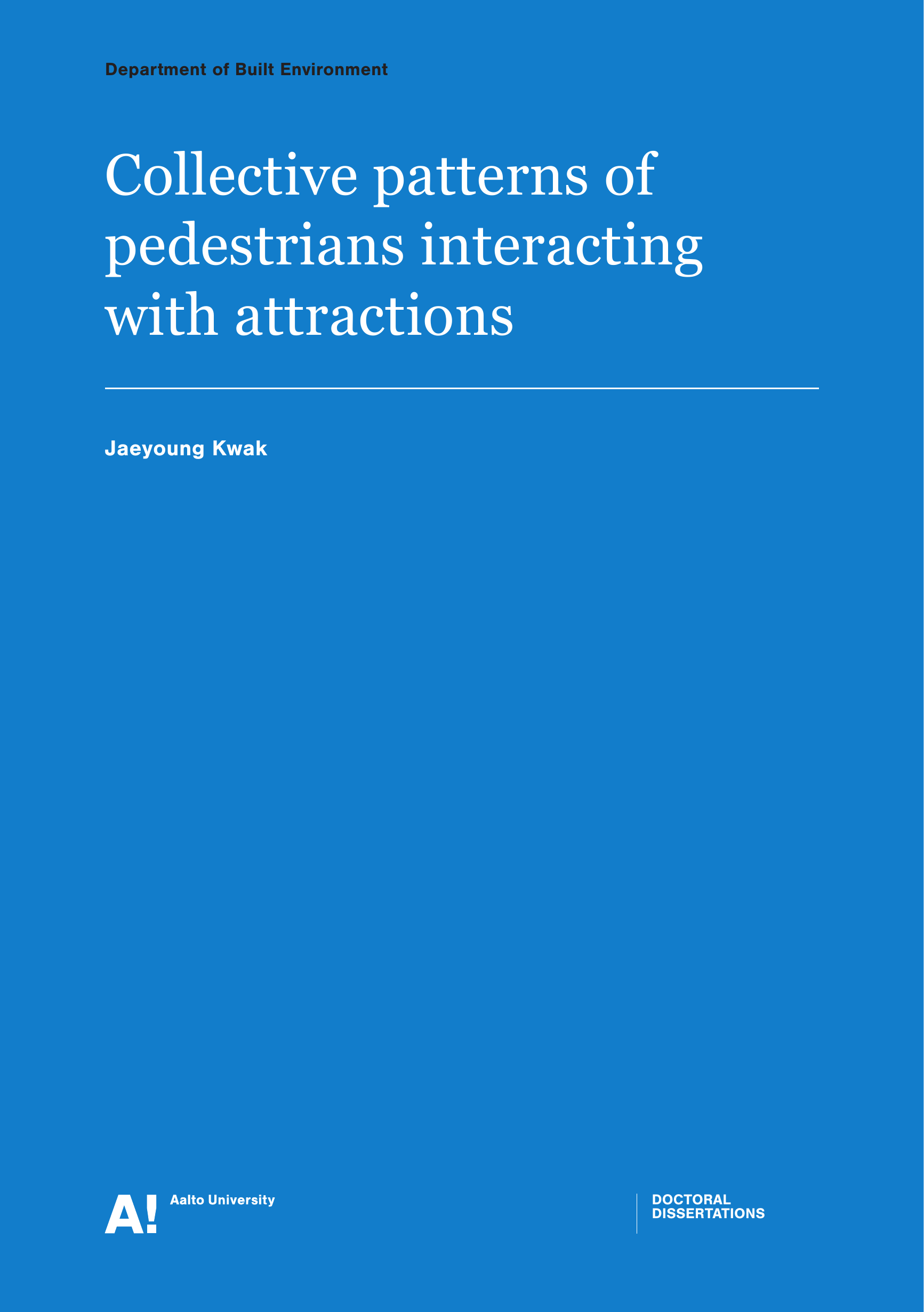}

\author{Jaeyoung Kwak}

\title{Collective patterns of pedestrians interacting with attractions}

%
\draftabstract{
Walking is a fundamental activity of a human life, not only for moving between places but also for interacting with surrounding environments. While walking to destinations, pedestrians may acquaint themselves with attractions such as artworks, shop displays, and public events. If such attractions are tempting enough, pedestrians opt to stop walking to join the attractions. Although the existence of attractions may considerably affect the pedestrian flow patterns, little attention has been paid to the interactions between pedestrians and attractions, and their impacts on pedestrian traffic.

This dissertation developed two microscopic models of pedestrian flow interacting with attractions. In numerical simulations, the presented models were examined by systematically controlling model parameters. After performing the numerical simulations, various collective patterns were identified and summarized in phase diagrams based on macroscopic measures. By doing so, this dissertation investigated the dynamics of pedestrian flow interacting with attractions from the perspective of attracted pedestrians and passersby.

The first model represents the attractive force towards the attractions in line with the social force models. The first model predicted various collective behaviors of attracted pedestrians, such as forming stable clusters around the attractions and rushing into the attractions. To understand collective patterns of pedestrians' visiting behavior, the second model was formulated as a function of the social influence and the average duration of visiting an attraction. The second model suggested that increasing the social influence and the average duration of visiting an attraction tended to attract more visitors to the attraction, but the increment was effective for a certain range of the parameters. Based on the second model, jamming transitions in pedestrian flow interacting with an attraction were also studied. It was found that an attendee cluster can trigger jamming transitions by increasing conflicts among pedestrians near the attraction.

This dissertation contributes to the current body of knowledge on the collective behavior of pedestrian motions. This is achieved by modeling pedestrian dynamics interacting with attractions and providing possible explanations of the collective patterns. The presented models and their applications enable one to understand various collective patterns of pedestrians interacting with attractions. The findings presented in this dissertation can provide an insight into pedestrian flow patterns in stores and improve the understanding of collective phenomena relevant to pedestrian facility management.
}
%

%
\begin{preface}
During my years at Aalto University, I have an enjoyable period of learning not only in the research arena but also on a personal level. Numerous individuals helped my learning both in-- and outside the academic sphere. I greatly appreciate their kind support. Without their support, this dissertation would not have come into existence. I would like to offer my special thanks to the people who supported and helped me so much throughout this period.

I would like to express my very great appreciation to my supervisor Professor Tapio Luttinen for patiently guiding this work and offering the freedom to explore the unknowns in the area of pedestrian flow dynamics. I wish to acknowledge the help provided by my thesis advisors Drs. Hang-Hyun Jo and Iisakki Kosonen. I am particularly grateful to Dr. Jo for helping me go through the research problems. I appreciate Dr. Kosonen for his commentary role in this work.

I would also like to thank preliminary examiners Professors Majid Sarvi and Daichi Yanagisawa for giving me valuable comments that helped me to improve this dissertation. Additionally, I would like to express my gratitude towards Professor Winne Daamen for being my opponent. 

I would like to acknowledge financial support from Aalto University 4D--Space MIDE project, School of Engineering Doctoral Program, and Energy Efficiency Research Program. CSC-IT Center for Science, Finland provided me computational resources.

Finally, I would like to thank my family for their unconditional love and support over the years.
\end{preface}

\tableofcontents


\listofpublications


%
\mbox{}
%
%
%

\nomenclature[A]{$\vec{f}_{i,d}$}{Driving force of pedestrian $i$}
\nomenclature[A]{$\vec{f}_{ij}$}{Repulsive force between pedestrian $i$ and pedestrian $j$}
\nomenclature[A]{$\vec{f}_{iB}$}{Repulsive force between pedestrian $i$ and boundary $B$}
\nomenclature[A]{$\vec{f}_{iA}$}{Attractive force between pedestrian $i$ and attraction $A$}
\nomenclature[A]{$\vec{g}_{ij}$}{Interpersonal elastic force between pedestrian $i$ and pedestrian $j$}
\nomenclature[A]{$h(\cdot)$}{Ramp function; $h(x)$ yields $x$ if $x > 0$, otherwise $0$}
\nomenclature[A]{$\vec{e}_i$}{Unit vector pointing desired walking direction of pedestrian $i$}
\nomenclature[A]{$\vec{e}_{i, 0}$}{Unit vector pointing initial desired walking direction of pedestrian $i$}
\nomenclature[A]{$\vec{e}_{ij}$}{Unit vector pointing from pedestrian $j$ to pedestrian $i$}
\nomenclature[A]{$\vec{e}_{iB}$}{Unit vector pointing from boundary $B$ to pedestrian $i$}
\nomenclature[A]{$\vec{e}_{iA}$}{Unit vector pointing from attraction $A$ to pedestrian $i$}
\nomenclature[A]{$\vec{t}_{ij}$}{Unit vector in shearing direction; perpendicular to $\vec{e}_{ij}$}
%
%
\nomenclature[A]{$C_x$}{Social force interaction strength; $x = p, b, r, s$ ($p$: interpersonal repulsion, $b$: boundary repulsion, $r$: repulsive interaction from attraction, $a$: attractive interaction from attraction)}
%
%
\nomenclature[A]{$l_x$}{Social force interaction range; $x = p, b, r, s$ ($p$: interpersonal repulsion, $b$: boundary repulsion, $r$: repulsive interaction from attraction, $a$: attractive interaction from attraction)}
\nomenclature[A]{$C$}{Relative attraction strength, i.e., $C = C_a/C_r$}
\nomenclature[A]{$k_x$}{Elastic constants; $x = n, t$ ($n$: normal, $t$: tangential)}
\nomenclature[A]{$\vec{d}_{ij}$}{Distance vector pointing from pedestrian $j$ to pedestrian $i$, i.e., $\vec d_{ij}\equiv \vec x_i-\vec x_j$}
\nomenclature[A]{$d_{ij}$}{Distance between pedestrian $i$ and pedestrian $j$, i.e., $d_{ij} = \left"\"| \vec d_{ij} \right"\"|$}
\nomenclature[A]{$d_{iB}$}{Distance between pedestrian $i$ and boundary $B$}
\nomenclature[A]{$d_{iA}$}{Distance between pedestrian $i$ and attraction $A$}
\nomenclature[A]{$b_{ij}$}{Effective distance between pedestrian $i$ and pedestrian $j$}
\nomenclature[A]{$v_d$}{Pedestrian desired speed}
\nomenclature[A]{$v_{\rm max}$}{Pedestrian maximum speed}
\nomenclature[A]{$\vec{v}_i$}{Velocity of pedestrian $i$}
\nomenclature[A]{$v_i$}{Speed of pedestrian $i$} 
\nomenclature[A]{$\vec{v}_j$}{Velocity of pedestrian $j$}
\nomenclature[A]{$v_j$}{Speed of pedestrian $j$} 
\nomenclature[A]{$\tau$}{Pedestrian relaxation time}
\nomenclature[A]{$\Delta t$}{Simulation time step}
\nomenclature[A]{$\Delta t_{s}$}{Stride time}
\nomenclature[A]{$y_{ij}$}{Relative displacement of pedestrian $i$ and pedestrian $j$ \\} 
\nomenclature[A]{$w_{ij}$}{Anisotropic function defined between pedestrian~$i$ and pedestrian~$j$}
\nomenclature[A]{$\lambda_{ij}$}{Pedestrian $i$'s minimum anisotropic strength against pedestrian $j$}
\nomenclature[A]{$\phi_{ij}$}{Angle between pedestrian $i$'s velocity vector and relative location of pedestrian $j$ with respect to pedestrian $i$}
\nomenclature[A]{$r_i$}{Radius of pedestrian $i$} 

\nomenclature[B]{$P_a$}{Joining probability}
\nomenclature[B]{$s$}{Social influence parameter}
\nomenclature[B]{$t_d$}{Average duration of visiting an attraction}
\nomenclature[B]{$N_a$}{The number of pedestrians who have already joined the attraction}
\nomenclature[B]{$N_0$}{The number of pedestrians who have not joined the attraction yet}
\nomenclature[B]{$K_x$}{Baseline values; $x = 0, a$ ($0$: for $N_0$, $a$: for $N_a$)}
\nomenclature[B]{$R_a$}{The range of perception}

\nomenclature[C]{$T_c$}{Time-to-collision}
\nomenclature[C]{$r_c(t)$}{Attendee cluster size at time $t$, i.e., $r_c = r_c(t)$}
\nomenclature[C]{$\psi$}{Streamline function of passerby flow}

\nomenclature[D]{$E$}{Efficiency of motion}
\nomenclature[D]{$E_i$}{Pedestrian $i$'s efficiency of motion}
\nomenclature[D]{$E(x, t)$}{Local efficiency of motion in a 1~m long segment $x$ at time $t$}
\nomenclature[D]{$E(x)$}{Stationary state average of local efficiency in a 1~m long segment~$x$, i.e., $E(x) = \langle E(x, t) \rangle$}
\nomenclature[D]{$E_a$}{Minimum value of $E(x)$ near the attraction}
\nomenclature[D]{$E_{up}$}{Minimum value of $E(x)$ upstream of the attraction}
\nomenclature[D]{$K$}{Normalized kinetic energy}

\nomenclature[D]{$N_v$}{The number of pedestrians near the attraction who have visited the attraction}
\nomenclature[D]{$N_p$}{The number of pedestrians near the attraction, i.e., within a range of $R_a$ from the attraction}
\nomenclature[D]{$N_p(x, t)$}{The set of passersby in a $1$~m long segment $x$ at time $t$}
\nomenclature[D]{$N_p(A, t)$}{The set of passersby in a section of $25\text{~m} \leq x \leq 35\text{~m}$ at time $t$}

\nomenclature[D]{$P_f$}{Freezing probability}
\nomenclature[D]{$N_{c, i}(t)$}{The number of conflicts experienced by passerby $i$ at time $t$}
\nomenclature[D]{$n_c(t)$}{Conflict index at time $t$}

\nomenclature[E]{$L$}{Corridor length}
\nomenclature[E]{$W$}{Corridor width}
\nomenclature[E]{$Q$}{Pedestrian influx}

\renewcommand{\nomname}{List of Frequently Used Symbols}
\printnomenclature

\chapter{Introduction}
\label{sec:intro}

\section{Background}
\label{sec:Background}
Collective dynamics of many-body systems has attracted much attention to traffic and related systems, such as granular particles~\cite{To_PRL2001, Zuriguel_PRL2011}, vehicles~\cite{Chowdhury_PhysRep2000, Helbing_RMP2001, Nagatani_RPP2002}, pedestrians~\cite{Helbing_Nature2000, Helbing_Nature1997}, and animals~\cite{Couzin_ProcB2002, John_PRL2009, Vicsek_PRL1995}. This subject has been studied by developing a set of individual behavioral rules in order to quantify emergent collective patterns from interactions among individuals. The beginning of such a bottom-up approach can be traced to the work of Maxwell and Boltzmann~\cite{Ball_2006, Cercignani_1998, Ruhla_1992}. Maxwell suggested the concept of the ideal gas in which every gas particle is subject to the same equation of motion. He developed a macroscopic description of gases based on microscopic descriptions of particle behavior. Later, Boltzmann proposed the law of increasing entropy by extending Maxwell's approach of gas dynamics. He stated that the entropy of a system, a macrostate often interpreted as a degree of disorder in a system of particles, grows as the number of possible microstates increases. Here, a microstate represents a configuration of an individual atom or molecule such as position and velocity. A macrostate is defined as an observable quantity of the system, for instance, the temperature or the number of gas particles in a container. The work of Maxwell and Boltzmann demonstrated how one can explain an observed macroscopic pattern in terms of individual particle behaviors. Still today, the line of thoughts presented by them is highly relevant to understand collective patterns of motions. 

Based on the bottom-up approach, various interesting collective behaviors have been identified in traffic flow studies. For example, Nagel and Schreckenberg explained traffic jam patterns based on their particle hopping model~\cite{Nagel_PRE1996, Nagel_JPhys1992}. In the model, a highway section is represented as a one-dimensional array of multiple cells and each cell either can be occupied by one vehicle or can be empty. The velocity of a vehicle is expressed in the number of cells hopping forward at the next time step. All the vehicles update their position based on four update rules: acceleration, slowing down, randomization, and car motion rules. According to the acceleration rule, a vehicle can proceed one more cell at the next time step if the distance to the vehicle ahead is large enough. The slowing down rule states that the vehicle reduces the number of cells hopping forward at the next time step based on the distance to the vehicle ahead. The randomization rule represents fluctuations in the vehicle motions by reducing the number of cells hopping forward at the next time step by one with a certain probability. After these three rules are applied to all vehicles, according to the car motion rule, the vehicles move ahead with the updated velocity. Nagel and Schreckenberg explained complex traffic flow patterns with a set of rules describing individual vehicle motions. 

Examples of the bottom-up approach can be also found in pedestrian flow studies. For instance, Helbing and his colleagues studied the formation of pedestrian trail systems based on an active walker model~\cite{Helbing_Nature1997, Helbing_PRE1997}. Their active walker model includes three sets of equations for pedestrian motion, ground condition, and pedestrian orientation. The equation of motion describes position and velocity of individual pedestrians. The equation of ground condition reflects creation and destruction of footprints. The equation of pedestrian orientation evaluates pedestrians' desired walking direction based on the ground condition. Their model explained how different topological structures of human trail systems form. It was demonstrated that formations of such trail systems can be described as self-organization phenomena. The self-organization phenomena have been one of the current central topics in pedestrian studies.

A considerable amount of literature has reported various interesting collective patterns of pedestrian motions such as lane formation~\cite{Burstedde_PhysicaA2001, Helbing_PRE1995, Yu_PRE2005}, faster-is-slower effect~\cite{Helbing_Nature2000}, and turbulent movement~\cite{Helbing_PRE2007, Ma_JSTAT2013, Moussaid_PNAS2011}. These collective patterns arise from repulsive interactions among pedestrians which can be observed when they are moving from one place to another. 

Not only moving between places, pedestrians are also interacting with surrounding environment especially attractions such as shop displays and public events. It has been well recognized that existence of such attractions can influence pedestrian flow patterns. For instance, Goffman~\cite{Goffman_1971} described that window shoppers act like obstructions to passersby on the streets when they stop to check store displays. Those shoppers can further interfere with other pedestrians when the shoppers enter and leave the stores. In another study, Helbing and Moln\'{a}r~\cite{Helbing_PRE1995} stated that attracted pedestrians form groups near attractions because of attractive interactions. Furthermore, one can infer that existence of the attractions in stores might be relevant to extreme pedestrian behaviors observed during shopping holidays such as Black Friday in the United States~\footnote{Over 20 video clips are available on YouTube.com showing people fighting over merchandise and many news articles have reported about Black Friday incidents. Examples include http://www.youtube.com/watch?v=K3RDTxVCKC4, http://www.independent.co.uk/news/world/americas/death-counter-records-number-of-people-killed-or-injured-on-black-friday-a6731766.html, and http://www.forbes.com/sites/adriankingsleyhughes/2012/11/24/black-friday-2012-shoppers-fight-over-cheap-deals}. Despite its relevance to pedestrian behavior, up to now, far too little attention has been paid to the collective dynamics of the attractive interactions between pedestrians and attractions. The effects of such attractive interactions will be the topic of this dissertation. 

\section{Research Questions}
\label{sec:RQ}
This dissertation aims at investigating collective patterns of pedestrians interacting with attractions by developing numerical simulation models. Three research questions are developed in order to address the effect of attractive interactions mentioned in Section~\ref{sec:Background}, such as pedestrian groups near attractions and their impact on pedestrian flow. 

The first research question is formulated as, \textbf{(RQ1) how do collective patterns of pedestrian motions emerge from the attractive interactions between pedestrians and attractions?} This research question is inspired by the impulse stops~\cite{Borgers_GeoAnal1986, Timmermans_TrB1992}. The impulse stop is a behavior of visiting an attraction without planning to do so in advance. An attraction is a place or an object that might be interesting for pedestrians, such as shop displays and public events. Understanding such impulse stops can support effective design and management of pedestrian facilities, for instance, placing merchandise in stores in order to encourage consumers to stop to buy~\cite{Hui_JMR2013}. However, little attention has been paid to the attractive interactions between pedestrians and attractions. 

It is widely believed that individual choice behavior of joining an attraction can be influenced not only by the attractiveness of the attractions but also by social influence from other individuals. For instance, previous studies on stimulus crowd effects reported that a pedestrian is more likely to shift one's attention towards the crowd as its size grows~\cite{Gallup_PNAS2012, Milgram_JPSP1969}. This belief is also generally accepted in the marketing area, which can be interpreted that having more visitors in a store can attract more pedestrians to the store~\cite{Bearden_JCR1989, Childers_JCR1992}. Based on that belief, marketing strategies have focused on increasing the duration of visiting a store and the strength of social interactions~\cite{Kaltcheva_JMkt2006}. Although the understanding of such a social influence on joining behavior has been widely accepted and practiced, the idea has not been incorporated into microscopic pedestrian behavior models. Therefore, the second research question is formulated as \textbf{(RQ2) how can social influence on one's choice behavior shape the collective patterns of pedestrians' visiting behavior?} 

\textbf{RQ1} and \textbf{RQ2} focus on modeling the attractive interactions between pedestrians and attractions. Based on \textbf{RQ1} and \textbf{RQ2}, Studies~\Rmnum{1} and \Rmnum{2} have reported various collective patterns emerging from the attractive interactions. Nevertheless, little is known about their influence on passerby traffic. Pedestrians facilities are usually designed not only for accommodating pedestrians who are interested in shopping and attending social occasions, but also for passersby who regularly commute and walk through the facilities. It is apparent that if a large attendee cluster exists near an attraction, passersby are forced to walk through the reduced available space. Consequently, the attendee cluster is acting as a pedestrian bottleneck for passersby. The flow through the bottleneck can show transitions from the free flow state to the jamming state and may end in gridlock. Up to now, most of the pedestrian bottleneck studies have been performed for static bottlenecks, meaning that the bottlenecks are at fixed locations and their size does not change over time~\cite{Hoogendoorn_TrSci2005, Seyfried_TrSci2009}. One might think that assuming an attendee cluster as a static bottleneck is enough to understand the jamming transitions induced by an attraction. However, the assumption of the static bottleneck cannot reflect the interactions among the attraction, attracted pedestrians, and passersby contributing to the onset of jamming transitions. Accordingly, the third research question is formulated as \textbf{(RQ3) how can pedestrian flow interacting with an attraction result in pedestrian jams?} An attendee cluster is conceptualized as a dynamic bottleneck in the sense that its size changes over time according to the joining behavior of attracted pedestrians. 

\section{Research Approach}
In order to address research questions discussed in Section~\ref{sec:RQ}, I have extended microscopic pedestrian models for the attractive interactions between pedestrians and attractions. In Study~\Rmnum{1}, the attractive force is appended to the social force model in order to incorporate the attractive interactions between pedestrians and attractions. In Study~\Rmnum{2}, I introduce a joining probability model which can reflect social influence from other pedestrians. In Study~\Rmnum{3}, the presented models are further extended for simulating jamming patterns of pedestrians interacting with an attraction.

The presented models are then examined by means of numerical simulations. Simulation parameters are systematically controlled in order to observe various collective patterns arising from interactions between pedestrians and attractions. I devise macroscopic measures in order to quantify various collective patterns and then summarize the results in phase diagrams. In Study~\Rmnum{1}, the macroscopic measures are introduced to quantify various attendee cluster formations. In Study~\Rmnum{2}, pedestrian visiting patterns are distinguished. In Study~\Rmnum{3}, different jam patterns near an attraction are identified. By doing so, this dissertation investigates the dynamics of pedestrian flow interacting with attractions from the perspective of the activity needs for attracted pedestrians and the mobility needs for passersby. 

\section{Dissertation Outline}
This dissertation is organized as follows. Chapter~\ref{sec:LiteratureReview} provides a review of existing literature that is closely linked with the topic of this dissertation. Chapter~\ref{sec:Models} describes presented numerical simulation models and their setups. The results of numerical simulations are presented in Chapter~\ref{sec:Results}. Finally, Chapter~\ref{sec:Conclusions} summarizes the main findings, discusses limitations, and suggests future research directions.

\chapter{Related Work}
\label{sec:LiteratureReview}
This chapter provides a review of existing literature that is closely linked with the topic of this dissertation. In Section~\ref{sec:review_modeling}, I provide an overview of the existing pedestrian models on microscopic and macroscopic scales. First, I describe macroscopic pedestrian flow models which have been developed based on fluid dynamic equations (Section~\ref{sec:macroscopic}). In macroscopic models, pedestrian streams are considered similar to fluid streams, so the large-scale pedestrian motion is the major interest. Then, I present microscopic pedestrian flow models including cellular automata (Section~\ref{sec:CA}), velocity-based models (Section~\ref{sec:velocity-based}), and force-based models (Section~\ref{sec:force-based}). In microscopic models, each pedestrian is treated like an individual particle, thus the modeling effort is focusing on developing mathematic expressions of individual motions. Section~\ref{sec:modeling_remarks} summarizes the presented modeling approaches and identifies research directions. Section~\ref{sec:Phases} discusses the concept of phases in pedestrian flow, which is important to quantify numerical simulation results presented in this dissertation. 

\section{Modeling Approaches}
\label{sec:review_modeling}

\subsection{Macroscopic Models}
\label{sec:macroscopic}
Continuity equation in fluid mechanics states that the mass within a finite control volume remains constant because of the rates of mass inflow and outflow balance out through the control volume~\cite{Anderson_2010, Batchelor_2000, Munson_2013}. The differential form of the continuity equation is
\begin{equation}
\label{eq:continuity}
	\frac{\partial \rho}{\partial t} + \nabla \rho \vec{v} = 0,
\end{equation}
where $\rho$ is density and $\vec{v}$ is velocity vector of a fluid stream. 

\subsubsection{Hughes models}
Hughes~\cite{Hughes_TrB2002} developed a macroscopic pedestrian flow model analogous to Lighthill–Whitham–Richard (LWR) macroscopic modeling approach in vehicular traffic flow studies~\cite{Lighthill_ProcA1955, Richards_OR1956}. First, he formulated that walking direction of a pedestrian stream is perpendicular to the potential $\Phi$:
\begin{equation}\label{eq:Hughes_1}
	\begin{split}
	\hat{\Phi}_x &= -\frac{\partial \Phi / \partial x }{\sqrt{(\partial \Phi / \partial x)^2+(\partial \Phi / \partial y)^2}}\\
	\hat{\Phi}_y &= -\frac{\partial \Phi / \partial y }{\sqrt{(\partial \Phi / \partial x)^2+(\partial \Phi / \partial y)^2}}.
	\end{split}	
\end{equation}
Here, $\hat{\Phi}_x$ and $\hat{\Phi}_y$ indicate directional cosines of pedestrian stream motion. Second, he represented the velocity of a pedestrian stream $\vec{v} = (v_x, v_y)$ with speed function $f = f(\rho)$ and directional cosines $\hat{\Phi}_x$ and $\hat{\Phi}_y$,
\begin{equation}\label{eq:Hughes_2}
	\begin{split}
	v_x &= f(\rho) \hat{\Phi}_x\\
	v_y &= f(\rho) \hat{\Phi}_y.
	\end{split}	
\end{equation}
Third, he also took into account that pedestrians tend to minimize their travel time while avoiding extremely high-density areas. He introduced walking strategy function,
\begin{equation}\label{eq:Hughes_3}
	\frac{1}{\sqrt{(\partial \Phi / \partial x)^2+(\partial \Phi / \partial y)^2}} =  g(\rho) \sqrt{v_x^2+v_y^2},
\end{equation}
where $g(\rho)$ is a discomfort function. Hughes~\cite{Hughes_TrB2002} set the value of $g(\rho)$ as 1 for most density levels but increased it for high pedestrian density situations. Based on the continuity equation in Equation~(\ref{eq:continuity}) and Equations~(\ref{eq:Hughes_1}) to (\ref{eq:Hughes_3}), Hughes presented the governing equations for pedestrian flow:
\begin{equation}\label{eq:macro_1}
	-\frac{\partial \rho}{\partial t} 
	+ \frac{\partial}{\partial x}\left( \rho g(\rho) f^2(\rho) \frac{\partial \Phi}{\partial x} \right)
	+ \frac{\partial}{\partial y}\left( \rho g(\rho) f^2(\rho) \frac{\partial \Phi}{\partial y} \right) = 0
\end{equation}
and 
\begin{equation}\label{eq:macro_2}
	\frac{1}{\sqrt{(\partial \Phi / \partial x)^2+(\partial \Phi / \partial y)^2}} =  g(\rho) f(\rho).
\end{equation}

Later, Huang~\textit{et al.}~\cite{Huang_TrB2009} interpreted the potential function $\Phi(x, y)$ as a walking cost potential. They assumed that a pedestrian stream follows a path minimizing walking cost according to a generalized cost function $C = C(\rho, x, y, t)$. In line with Eikonal equation~\cite{Jeong_SIAM2008, Sethian_SIAM1999, Zhang_JSC2006, Zhao_MathCompu2005}, they reformulated a walking strategy function of Hughes's model [Equation~(\ref{eq:Hughes_3})]:
\begin{equation}
	C \frac{\vec{q}}{\left\| \vec{q} \right\|} + \nabla \Phi = 0,
\end{equation}
where, the flow $\vec{q}$ is defined as a product of density and velocity, i.e., $\vec{q} = \rho \vec{v}$. 

\subsection{Cellular Automata}
\label{sec:CA}
In cellular automata (CA), pedestrian walking space is divided into uniform grids (or cells) and pedestrian movements are updated at discrete time step. Each pedestrian occupies a particular cell and moves between cells according to the prescribed rules. A pedestrian can move to a neighboring cell and the set of neighboring cells can be defined in two ways: von Neumann neighborhood and Moore neighborhood. In a rectangular grid, the von Neumann neighborhood includes a central cell and its four adjacent cells neighboring by edges of the central cell. The Moore neighborhood consists of a central cell and the eight cells surrounding the central one, adding diagonal connectivity to the von Neumann neighborhood. The CA-based approaches have been one of the popular pedestrian flow models because the approaches offer simple and efficient computation.

\subsubsection{Gipps-Marksj\"{o} model}
Gipps and Marksj\"{o}~\cite{Gipps_MCS1985} proposed the gain-cost model in which the walking space is discretized into $0.5~\text{m} \times 0.5~\text{m}$ square cells. Each cell can be occupied by no more than one pedestrian. They modeled that a pedestrian in cell $i$ moves to cell $j$ which maximizes the benefit of his movement. Similar to the Moore neighborhood, a set of available cells includes cell $i$ occupied by the pedestrian and eight neighboring cells. The benefit $B_j$ for cell $j$ is calculated by subtracting cost score $S_j$ from gain function value $P(\sigma_j)$, i.e., 
\begin{equation}
	B_j = P(\sigma_j) - S_j.
\end{equation}
The cost score $S_j$ represents the repulsion effects from nearby pedestrians and obstacles assigned to the cell $j$. The gain function $P(\sigma_j)$ reflects the distance to the destination $d$. 
The cost score function $S_j$ is given as 
\begin{equation}
	S_j = \frac{1}{(\Delta - \alpha)^2 + \beta},
\end{equation}
where $\Delta$ is the distance between cell $j$ and a cell $i$ in which the pedestrian is standing. Here, $\alpha = 0.4$ is a constant which is slightly smaller than the size of a pedestrian ($= 0.5$~m), and $\beta = 0.015$ is an arbitrary constant which moderates fluctuation in the cost score function $S_j$. The gain function $P(\sigma_j)$ is given as 
\begin{equation}
	P(\sigma_j) = K \left| \cos \sigma_j \right| \cos \sigma_j,
\end{equation}
where $K$ is gain function constant and $\sigma_j$ is the angle between a vector pointing from current position to a destination and a vector pointing to cell $j$ where the pedestrian plans to move. If the pedestrian is not going to move, $P(\sigma_j)$ becomes zero. According to the definition of a dot product between two vectors, $\cos \sigma_j$ is calculated as 
\begin{equation}
	\cos \sigma_j = \frac{(\vec{x}_j-\vec{x}_i) \cdot (\vec{x}_d-\vec{x}_i)}{\left| \vec{x}_j-\vec{x}_i \right| \left| \vec{x}_d-\vec{x}_i \right|},
\end{equation}
where $\vec{x}_j$, $\vec{x}_i$, and $\vec{x}_d$ are the location of cell $j$, cell $i$, and the destination $d$, respectively. 

\subsubsection{Blue-Adler model}
Similar to the particle hopping model of Nagel and Schreckenberg~\cite{Nagel_PRE1996, Nagel_JPhys1992}, Blue and Alder~\cite{Blue_TRR1998, Blue_TRR1999, Blue_TrB2001} suggested an alternative formulation of cellular automata for pedestrian motions based on three sets of behavioral rules including sidestepping, forward movement, and conflict mitigation. Sidestepping rules describe lateral movements of pedestrians: move to left or right lanes, or stay in the same lane. Forward movement rules decide the desired speed and available gap ahead. Conflict mitigations enable pedestrians to avoid a head-on collision between approaching pedestrians from opposite directions. Pedestrian movements are updated with the probability of movement related to the behavioral rules. 

\subsubsection{Floor field model}
Burstedde~\textit{et al.}~\cite{Burstedde_PhysicaA2001} and Kirchner and Schadschneider~\cite{Kirchner_PhysicaA2002} introduced the floor field model. In the floor field model, pedestrians are walking on a floor field which is represented as a superposition of a static floor field and a dynamic floor field. The static floor field indicates quantities fixed over time, such as the shortest distance to an exit. The dynamics field represents virtual traces created by pedestrians who passed a certain area. The concept of dynamic floor field is analogous to chemotaxis of ants. In ant traffic systems, leading ants secrete pheromones when they march towards a food source, so following ants can reach the source by chemotaxis. The concept of floor field has been applied to develop pedestrian navigation models, such as in studies performed by Asano~\textit{et al.}~\cite{Asano_TrC2010}, Hartmann~\cite{Hartmann_NJP2010}, and Kneidl~\textit{et al.}~\cite{Kneidl_TrC2013}. 

The floor field model describes pedestrian movements by means of a transition probability. The transition probability $p_{ij}$ for moving to a neighboring cell $(i, j)$ is
\begin{equation}
	p_{ij} = N \exp(k_D D_{ij} + k_S S_{ij}) (1-n_{ij}) \xi_{ij},
\end{equation}
where $D_{ij}$ and $S_{ij}$ are strength of dynamic and static fields at cell $(i, j)$, and $k_D$ and $k_S$ are sensitivity parameters of $D_{ij}$ and $S_{ij}$, respectively. Occupation number $n_{ij}$ is 1 if cell $(i, j)$ is occupied by a pedestrian, 0 for otherwise. Obstacle number $\xi_{ij}$ becomes 0 for a cell obstacles or walls, while it is 1 for cells where pedestrians can move to. Normalization factor $N$ is given as
\begin{equation}
	N = \frac{1}{\sum_{(i, j)}^{ } \exp(k_D D_{ij} + k_S S_{ij}) (1-n_{ij}) \xi_{ij}}.
\end{equation}
After calculating the transition probability $p_{ij}$, each pedestrian decides where he or she will move and then updates one's position. 

\subsection{Velocity-based Models}
\label{sec:velocity-based}
In velocity-based models, pedestrian motions are described by a first-order ordinary differential equation,
\begin{equation}
	\frac{d\vec{x}_i(t)}{dt} = \vec{v}_i(t).
\end{equation}
Here, $\vec{x}_i(t)$ is the position of pedestrian $i$ at time $t$ and $\vec{v}_i(t)$ is his velocity at time $t$. In contrast to CA-based approaches, the velocity-based models can express pedestrian position and speed as continuous variables. The velocity-based models can simulate pedestrian movements without calculating force terms, so inertia effects do not occur. Thus, the implementation is easier than force-based models which are described by a second-order ordinary differential equation (see Section~\ref{sec:force-based}). Different velocity-based models have been proposed by introducing various formulations of $\vec{v}_i(t)$.

\subsubsection{Contractile particle model}
Baglietto and Parisi~\cite{Baglietto_PRE2011} presented the contractile particle model in which pedestrians are represented as circles with variable radii. Velocity of pedestrian $i$ is described with desired velocity $\vec{v}_{i, d}$ and escape velocity $\vec{v}_{i, e}$. 
\begin{equation}
	\vec{v}_i(t) = \left\{ \begin{tabular}{ll}
	$\vec{v}_{i, d}$ & if pedestrian $i$ is free of contact,\\
	$\vec{v}_{i, e}$ & otherwise,
	\end{tabular}\right.
\end{equation}
where $\vec{v}_{i, d}$ is the desired velocity pointing destination from current position $\vec{x}_i$, and $\vec{v}_{i, e}$ is the escape velocity from the boundary of obstacles or other pedestrians in contact. The desired velocity $\vec{v}_{i, d}$ is
\begin{equation}
	\vec{v}_{i, d} = v_{i, d} \vec{e}_i,
\end{equation}
where $v_{i, d}$ is the desired speed and $\vec{e}_i$ is a unit vector in the desired direction. The desired speed is formulated as
\begin{equation}\label{eq:vd_contractile}
	v_{i, d} = v_{d, \max} \left( \frac{r_{i} - r_{\min}}{r_{\max} - r_{\min}} \right)^{\beta},
\end{equation}
where $v_{d, \max}$ is the maximum desired speed which a pedestrian can attain if his movement is not restricted by other pedestrians or obstacles. The pedestrian radius $r_i$ varies between the maximum radius $r_{\max}$ and the minimum radius $r_{\min}$. The exponent $\beta$ controls the shape of desired speed function in Equation~(\ref{eq:vd_contractile}). 

The escape velocity $\vec{v}_{i, e}$ is given as
\begin{equation}
	\vec{v}_{i, e} = v_e \frac{\sum_{j}^{ }\vec{e}_{ij}}{\left\| \sum_{j}^{ } \vec{e}_{ij} \right\|},
\end{equation}
where $v_e$ is the escape speed and $\vec{e}_{ij}$ is a unit vector pointing from pedestrian $j$ to $i$. The escape speed $v_e$ set to be the same as the maximum value of desired speed, i.e., $v_e = v_{d, \max}$.

\subsubsection{Gradient navigation model}
Dietrich and K\"{o}ster~\cite{Dietrich_PRE2014} introduced the gradient navigation model. In the model, pedestrian motions are described with relaxed speed $w_i(t)$ and navigation function $\vec{N}(\vec{x}_i, t)$: 
\begin{equation}
	\frac{\mathrm{d} \vec{x}_i(t)}{\mathrm{d} t} = w_i(t) \vec{N}(\vec{x}_i, t)
\end{equation}
Here, the navigation function $\vec{N}(\vec{x}_i, t)$ represents desired walking direction. The relaxed speed $w_i(t)$ is given as
\begin{equation}
	\frac{\mathrm{d} w_i(t)}{\mathrm{d} t} = \frac{1}{\tau} \left[ v_d \left\| \vec{N}(\vec{x}_i, t) \right\| -w_i(t) \right],
\end{equation}
where $\tau$ is relaxation constant, and $v_d$ is the desired speed of pedestrian $i$ which is determined based on the local crowd density $\rho(x_i)$. The navigation function $\vec{N}$ is given as a superposition of navigation vectors $\vec{N}_T$ and $\vec{N}_P$, and these navigation vectors are given as gradient of distance functions. i.e., 
\begin{equation}
	\vec{N}(\vec{x}_i, t) = g(g(\vec{N}_T)+g(\vec{N}_P)),
\end{equation}
where, $g(\vec{x})$ is a scaling function normalizing a vector $\vec{x}$ to a length in $[0, 1]$. Here, $\vec{N}_T$ is desired walking direction vector which minimizes walking time to destination $\sigma$, 
\begin{equation}
	\vec{N}_T = -\nabla \sigma.
\end{equation}
Repulsion effects from other pedestrians and obstacles are represented by $\vec{N}_P$,
\begin{equation}
	\vec{N}_P = -\left( \sum_{j\neq i}^{ } \nabla P_{i,j} + \sum_{B}^{ } \nabla P_{i,B} \right),
\end{equation}
where $P_{i,j}$ and $P_{i,B}$ are functions of distance to other pedestrian $j$ and obstacle $B$, respectively. See Dietrich and K\"{o}ster~\cite{Dietrich_PRE2014} for further details of the model. 

\subsection{Force-based Models}
\label{sec:force-based}
While velocity-based models are formulated based on a first-order ordinary differential equation, force-based models predict pedestrian motions with a second-order ordinary differential equation, i.e., 
\begin{equation}
	\frac{\mathrm{d}^2 \vec{x}_i(t)}{\mathrm{d} {t}^2} = \vec{f}_i(t).
\end{equation}
Here, $\vec{f}_i(t)$ is the force acting on pedestrian $i$ at time at time $t$. The force-based models represent interactions among pedestrians with virtual forces and can consider physical contacts among pedestrians similar to the case of granular flows~\cite{Duran_1999}. Different force-based models have been proposed based on analogies with various forces in Newtonian mechanics. The force-based models have been developed in order to describe pedestrian behaviors in detail, especially for the case of high pedestrian density. 

\subsubsection{Magnetic force model}
By the analogy with Coulomb's law, Okazaki~\cite{Okazaki_1979} proposed the magnetic force model. In the model, each pedestrian is modeled as a positively charged particle, and similarly, obstacles such as walls and columns are also modeled as positive poles. In contrast, the destination points are represented as negative poles. According to the magnetic force model, pedestrian $i$ walking towards his destination is described as a positively charged particle attracted by a negative pole. Avoiding other pedestrians and obstacles are described by repulsive accelerations acting on pedestrian $i$ from positively charged particles and poles. The magnetic force $\vec{F}_{i, m}$ exerts on pedestrian $i$ from a negative pole is given as
\begin{equation}
	\vec{F}_{i, m} = \frac{kq_{1}q_{2}\vec{r}}{r^3},
\end{equation}
where $k$ is magnetic force constant. The intensity of positively charged particle (i.e., pedestrian) is indicated by $q_1$, and that of negative pole (i.e., destination) is denoted by $q_2$. The distance vector from pedestrian $i$ to a destination is denoted by $\vec{r}$ and $r$ is its magnitude. The repulsive acceleration acting on pedestrian $i$ from pedestrian $j$ is given as 
\begin{equation}
	\vec{a}_{ij} = \vec{v}_i \cos(\alpha) \tan(\beta),
\end{equation}
where $\vec{v}_i$ is the velocity of pedestrian $i$. The angle $\alpha$ is defined between $v_i$ and $v_{ij}$. The angle $\beta$ is defined between $v_{ij}$ and a tangential line to the circumference of pedestrian $j$ from the position of pedestrian $i$. Here, $v_{ij}$ is the relative velocity of pedestrian $i$ to pedestrian $j$. Note that, the tangential line is crossing the velocity vector of pedestrian $j$.

\subsubsection{Social force models}
\label{sec:SFM}
Although the magnetic force model is considered as the first force-based model, the model has not been widely applied. This seems to be because the formulation and implementation of the magnetic force model are complicated and not straightforward. Helbing and Moln\'{a}r~\cite{Helbing_PRE1995} proposed the social force model, analogously to self-propelled particle models~\cite{DOrsogna_PRL2006, Mogilner_MathBio2003, Silverberg_PRL2013, Vicsek_PRL1995}. The social force model~\cite{Helbing_PRE1995} describes the pedestrian movements as a superposition of driving, repulsive, and attractive force terms. Each pedestrian $i$ is modeled as a circle with radius $r_i$ in a two-dimensional space. The position and velocity of each pedestrian $i$ at time $t$, denoted by $\vec{x}_i(t)$ and $\vec{v}_i(t)$, evolve according to the following equations:
\begin{equation}
	\frac{d\vec{x}_i(t)}{dt} =\vec v_i(t)
\end{equation}
and
\begin{eqnarray}\label{eq:EoM}
	\frac{d\vec{v}_i(t)}{dt} = \vec{f}_{i,d}+\sum_{j\neq i}^{ }{\vec{f}_{ij}}+\sum_{B}^{ }{\vec{f}_{iB}}+\sum_{A}^{ }{\vec{f}_{iA}}.
\end{eqnarray}
Here, the driving force $\vec{f}_{i,d}$ describes the pedestrian $i$ accelerating to reach its destination. The repulsive force between pedestrians $i$ and $j$, $\vec{f}_{ij}$, denotes the tendency of pedestrians to keep a certain distance from each other. The repulsive force from boundary $B$, $\vec{f}_{iB}$, shows the interaction between pedestrian $i$ and boundary $B$ (i.e., walls and obstacles). The attractive force $\vec{f}_{iA}$ indicates pedestrian movements toward attractive stimuli. The attractive stimuli can be, for instance, attractive interactions among pedestrian group members or between pedestrians and attractions such as shop displays and museum exhibits.

The driving force $\vec{f}_{i,d}$ is given as
\begin{eqnarray}\label{eq:driving}
	\vec{f}_{i,d} = \frac{v_d\vec{e}_{i}-\vec{v}_{i}(t)}{\tau},
\end{eqnarray}
where ${v}_{d}$ is the desired speed and $\vec{e}_{i}$ is a unit vector in the desired direction, independent of the position of the pedestrian $i$. The relaxation time $\tau$ controls how fast the pedestrian $i$ adapts its velocity to the desired velocity. 

The repulsive force between pedestrians $i$ and $j$, denoted by $\vec{f}_{ij}$, is the sum of the gradient of repulsive potential, 
\begin{eqnarray}\label{eq:f_ij}
	\vec{f}_{ij} &=& -\nabla_{\vec{d}_{ij}}V(b_{ij}).
\end{eqnarray}
Here, the repulsion potential represents the level of interpersonal repulsion associated with the relative positions of pedestrians $i$ and $j$. Similar to the relationship between force and potential energy in physics, the interpersonal repulsive force can be expressed as the negative of the derivate of the potential function. The repulsive potential is given as 
\begin{eqnarray}\label{eq:RepulsivePotential}
	V(d_{ij}) &=& C_p l_p\exp\left(-\frac{b_{ij}}{l_p}\right).
\end{eqnarray}
Here, $C_p$ and $l_p$ denote the strength and the range of repulsive interaction between pedestrians, and $b_{ij}$ is the effective distance between pedestrians $i$ and $j$. The social force models can be categorized according to the formulation of the effective distance $b_{ij}$. 

Circular specification (CS)~\cite{Helbing_RMP2001} is the most simplistic form of the repulsive interaction, assuming that $b_{ij}$ is as a function of $d_{ij}$, the distance between pedestrians $i$ and $j$,
\begin{equation}\label{eq:bij_CS}
	b_{ij} = d_{ij}-(r_i+r_j).
\end{equation}
The explicit form of $f_{ij}$ can be written as 
\begin{eqnarray}\label{eq:CS}
	\vec{f}_{ij} &=& C_p exp \left(\frac{r_i+r_j-d_{ij}}{l_p}\right) \vec{e}_{ij},
\end{eqnarray}
where $\vec{e}_{ij} = \vec d_{ij}/d_{ij}$ is a unit vector pointing from pedestrian $j$ to pedestrian $i$, and $\vec d_{ij}\equiv \vec x_i-\vec x_j$ is the distance vector pointing from pedestrian~$j$ to pedestrian~$i$. Due to its simplicity, CS has been widely applied for agent-based modeling~\cite{Helbing_Nature2000, Oliveira_PRX2016, Parisi_PhysicaA2009} and multiscale modeling~\cite{Corbetta_Dissertation2016, Hoogendoorn_TrC2015, Hoogendoorn_PhysicaA2014}. Although CS describes the pedestrian motion with a minimal number of parameters, it has limitation on reflecting the influence of relative velocity. 

Helbing and Moln\'{a}r~\cite{Helbing_PRE1995} introduced Elliptical specification (ES-1), considering pedestrian $j$'s stride, $\vec{y}_{ij}\equiv\vec{v}_{j}\Delta t_{s}$ with the stride time $\Delta t_{s}$. The effective distance $b_{ij}$ is given as 
\begin{equation}\label{eq:bij_ES1}
	\begin{split}
	b_{ij} &= \frac{1}{2}\sqrt{(\|\vec{d}_{ij}\|+\|\vec{d}_{ij}-\vec{v}_{j}\Delta t_{s}\|)^{2}-\|\vec{v}_{j}\Delta t_{s}\|^{2}} \\
	&= \frac{1}{2}\sqrt{(\|\vec{d}_{ij}\|+\|\vec{d}_{ij}-\vec{y}_{ij}\|)^{2}-\|\vec{y}_{ij}\|^{2}}.
	\end{split}
\end{equation}
Based on Equation~(\ref{eq:bij_ES1}), the explicit form of $\vec{f}_{ij}$ is given as
\begin{eqnarray}
	\vec{f}_{ij} &=& C_p\exp\left(-\frac{b_{ij}}{l_p}\right)\frac{\|\vec{d}_{ij}\|+\|\vec{d}_{ij}-\vec{y}_{ij}\|}{4b_{ij}}\left(\frac{\vec{d}_{ij}}{\|\vec{d}_{ij}\|}+\frac{\vec{d}_{ij}-\vec{y}_{ij}}{\|\vec{d}_{ij}-\vec{y}_{ij}\|}\right).\notag \\
	&\label{eq:ES1_explicit0}
\end{eqnarray}

Later, Johansson~\textit{et al.}~\cite{Johansson_ACS2007} introduced ES-2, an improved version of the ES-1, by revising $b_{ij}$ with relative displacement of pedestrians $i$ and $j$, $\vec{y}_{ij}\equiv (\vec{v}_{j}-\vec{v}_{i})\Delta t_{s}$. 
\begin{equation}\label{eq:ES2_1}
	\begin{split}
	b_{ij} &= \frac{1}{2} \sqrt{(\|\vec{d}_{ij}\| + \|\vec{d}_{ij}-(\vec{v}_{j}-\vec{v}_{i})\Delta t_{s}\|)^{2} - \|(\vec{v}_{j}-\vec{v}_{i})\Delta t_{s}\|^{2}} \\
	&= \frac{1}{2}\sqrt{(\|\vec{d}_{ij}\| + \|\vec{d}_{ij}-\vec{y}_{ij}\|)^{2}-\|\vec{y}_{ij}\|^{2}},
	\end{split}
\end{equation}

The interpersonal elastic force $\vec g_{ij}$ can be added to the interpersonal repulsion force term $\vec{f}_{ij}$ when the distance $d_{ij}$ is smaller than the sum $r_{ij} = r_{i}+r_{j}$ of their radii $r_{i}$ and $r_{j}$. Similar to the case of granular particles~\cite{Duran_1999}, Helbing~\cite{Helbing_RMP2001} suggested the interpersonal elastic force as:
\begin{eqnarray}\label{eq:friction_full}
	\vec{g}_{ij} = h(r_{ij}-d_{ij}) \left\{k_n \vec{e}_{ij}+k_t [(\vec{v}_j-\vec v_i)\cdot \vec{t}_{ij}]\vec{t}_{ij}\right\},
\end{eqnarray}
where $k_n$ and $k_t$ are the normal and tangential elastic constants. A unit vector $\vec{e}_{ij}$ is pointing from pedestrian $j$ to pedestrian $i$, and $\vec{t}_{ij}$ is a unit vector perpendicular to $\vec{e}_{ij}$. The function $h(x)$ yields $x$ if $x > 0$, while it gives $0$ if $x \leq 0$. Later, Moussa\"{i}d~\textit{et al.}~\cite{Moussaid_PNAS2011} presented a simpler form of the interpersonal compression mainly considering normal elastic force:
\begin{eqnarray}\label{eq:friction_simple}
	\vec{g}_{ij} = k_n h(r_{ij}-d_{ij}) \vec{e}_{ij}.
\end{eqnarray}

The repulsive force from boundaries is 
\begin{eqnarray}\label{eq:boundary}
	\vec{f}_{iB} = C_b\exp\left(-\frac{d_{iB}}{l_{b}}\right)\vec{e}_{iB},
\end{eqnarray}
where $d_{iB}$ is the perpendicular distance between pedestrian $i$ and wall, and $\vec{e}_{iB}$ is the unit vector pointing from the wall $B$ to the pedestrian $i$. The strength and the range of repulsive interaction from boundaries are denoted by $C_b$ and $l_b$.

In addition to the repulsive interactions, Helbing and Moln\'{a}r~\cite{Helbing_PRE1995} introduced the attractive force term, indicating attractive interactions between pedestrian group members and with attractions such as shop displays and museum exhibits. Later, Xu and Duh~\cite{Xu_IEEE2010} formulated the interaction between group members $i$ and $k$, $\vec{f}_{ik}$, as a summation of a bonding and a repulsive force between the members:
\begin{eqnarray}\label{eq:bonding}
	\vec{f}_{ik} = \left[ -C_b\exp\left(\frac{d_{ik}-r_{ik}}{l_{b}}\right) + \frac{C_p}{2}\exp\left(\frac{r_{ik}-d_{ik}}{l_{p}}\right) \right] \vec{e}_{ik}.
\end{eqnarray}
The first term on the right-hand side indicates the bonding effect between pedestrian $i$ and group member $k$, while the second term indicates the repulsive interaction between the members. The strength and the range of bonding force between pedestrians are denoted by $C_b$ and $l_b$, respectively. Note that $C_p$ is the strength of interpersonal repulsion between individual pedestrians who are not in the same pedestrian group. Here, $\vec{e}_{ik}$ is the unit vector pointing from group member $k$ to the pedestrian $i$. Xu and Duh~\cite{Xu_IEEE2010} set the strength of repulsive force between group members as $C_p/2$, indicating that the repulsive force strength between same group members is half of that among individuals who are not in the same group.

In another study, Moussa\"{i}d \textit{et al.}~\cite{Moussaid_PLOS2010} suggested a model of attractive interaction among pedestrian group members,
\begin{eqnarray}\label{eq:group}
	\vec{f}_{iG} = -\beta_{1}\alpha_{i}\vec{v}_i 
	+ q_{A}\beta_{2}\vec{e}_{Gi} + \sum_{k\neq i}^{ }{q_R \beta_3 \vec{e}_{ki}}.
\end{eqnarray}
The first term on the right-hand side reflects the gazing behavior of a group member $i$. The second term indicates the attractive force between group member $i$ and the center of the pedestrian group $G$. The third term shows the repulsive force between pedestrian $i$ group member $k$. Here, $\beta_1$ and $\beta_2$ are the strength of the attractive interactions between group members and for pedestrian $i$ and the group center $G$, respectively. The repulsive force strength $\beta_3$ is defined between pedestrian $i$ and group member $k$. Unit vectors $\vec{e}_{Gi}$ and $e_{ki}$ are pointing from pedestrian $i$ to the group center $G$ and pointing from pedestrian $i$ to group member $k$, respectively. The head rotation angle is denoted by $\alpha_1$. In addition, $q_A$ and $q_R$ are relevant to threshold distance between pedestrian $i$ and the group center $G$, and between pedestrian $i$ to group member $k$. If $q_A = 1$ and $q_R = 1$, the corresponding distance is smaller than the threshold values, otherwise 0.

\subsubsection{Centrifugal force models}
Yu~\textit{et al.}~\cite{Yu_PRE2005} proposed the centrifugal force model in which the repulsive force term is analogous to centrifugal force in mechanics. In the social force models, the repulsive force term is formulated as a derivative of repulsive potential. Although the social force models have been extended in order to incorporate relative velocity, the relative velocity is only associated with the interpersonal distance but not explicitly with the repulsive force term. In the centrifugal force model, the repulsive force term is formulated as a function of the relative velocity.
\begin{eqnarray}\label{eq:fij_CFM}
	\vec{f}_{ij} = -K_{ij}\frac{v_{ij}^2}{d_{ij}}\vec{e}_{ij},
\end{eqnarray}
where $v_{ij}$ is the projection of the relative velocity of pedestrians $i$ and $j$ in the direction $\vec{e}_{ij}$, 
\begin{eqnarray}
	\vec{v}_{ij} = \frac{1}{2} \left[ (\vec{v}_i-\vec{v}_j)\cdot \vec{e}_{ij} + \left\| (\vec{v}_i-\vec{v}_j)\cdot \vec{e}_{ij} \right\| \right].
\end{eqnarray}
An anisotropic effect factor $K_{ij}$ reflects that pedestrians react to others in front of them within their angle of view $180^\circ$,
\begin{eqnarray}
	{K}_{ij} = \frac{1}{2} \left[ \frac{\vec{v}_i \cdot \vec{e}_{ij} + \left\| \vec{v}_i \cdot \vec{e}_{ij} \right\|}{\left\| \vec{v}_i \right\|} \right].
\end{eqnarray}
Likewise, the boundary repulsive force is given in the centrifugal force model as
\begin{eqnarray}\label{eq:fiB_CFM}
	\vec{f}_{iB} = -K_{iB} \left( \frac{v_{iB}^2}{d_{ij}} \right) \vec{e_{iB}}
\end{eqnarray}
with
\begin{eqnarray}
	{K}_{iB} = \frac{1}{2} \left[ \frac{\vec{v}_i \cdot \vec{e}_{iB} + \left\| \vec{v}_i \cdot \vec{e}_{iB} \right\|}{\left\| \vec{v}_i \right\|} \right].
\end{eqnarray}

Later, Chraibi~\textit{et al.}~\cite{Chraibi_PRE2010} introduced the generalized centrifugal force model with consideration of desired speed for the repulsive force terms. The interpersonal repulsive force term $\vec{f_{ij}}$ is given as 
\begin{eqnarray}\label{eq:fij_GCFM}
	\vec{f}_{ij} = -K_{ij}\frac{(\eta v_d +v_{ij})^2}{d_{ij} - r_i(v_i) - r_j(v_j)}\vec{e}_{ij}.
\end{eqnarray}
Here, $\eta$ is the repulsive force strength parameter and $r_{i}$ is required space for pedestrian $i$'s motion. They suggested the required space for stride $r_i$ as a function of torso size and pedestrian $i$'s velocity,
\begin{eqnarray}\label{eq:ri_effective}
	{r}_{i}(v_i) = r_0 + v_i \Delta t_{s},
\end{eqnarray}
where $r_0$ is torso size and $\Delta t_{s}$ is stride time. Similar to Equation~(\ref{eq:fij_GCFM}), the boundary repulsive force is given as
\begin{eqnarray}\label{eq:fiB_GCFM}
	\vec{f}_{iB} = -K_{iB} \frac{(\eta v_d +v_{ij})^2}{d_{iB} - r_i(v_i)} \vec{e}_{iB},
\end{eqnarray}
where $b_{iB}$ is the distance between boundary $B$ and pedestrian $i$.

\subsection{Remarks}
\label{sec:modeling_remarks}

\subsubsection{Summary}
In Section~\ref{sec:review_modeling}, I have reviewed modeling approaches including macroscopic and microscopic approaches. From a viewpoint of fluid dynamics, macroscopic models treat pedestrians motions similar to fluid streams~\cite{Huang_TrB2009, Hughes_TrB2002}. This type of approach can directly estimate pedestrian flow characteristics such as pedestrian density and flow over a large area. However, due to its modeling approach, macroscopic models do not explicitly reflect interactions among pedestrians. Thus, macroscopic models are often limited to aggregated descriptions of pedestrian flow.

On the other hand, microscopic models view pedestrians as self-driven particles. Numerous approaches have been proposed such as cellular automata (CA)~\cite{Blue_TrB2001, Burstedde_PhysicaA2001, Gipps_MCS1985}, velocity-based models~\cite{Baglietto_PRE2011, Dietrich_PRE2014}, and force-based models~\cite{Helbing_PRE1995, Okazaki_1979, Yu_PRE2005}. In CA-based approaches, space is discretized into two-dimensional lattices where each cell can have at most one pedestrian. Pedestrians move between cells according to the probability of movement. The CA-based approaches offer simple and efficient computation but yield limited accuracy on describing pedestrian movements due to the nature of discretized time and space. 

Velocity-based models predict pedestrian movements based on a first-order ordinary differential equation. The velocity-based models can simulate pedestrian movements without calculating force terms, so inertia effects do not occur. However, the velocity-based models provide a limited representation of high-density situations when physical interactions among pedestrians become critical. 

Force-based models describe pedestrian motions as a combination of force terms that can be expressed as a second-order ordinary differential equation. Internal motivation and external stimuli are modeled as force terms, and pedestrian trajectories are calculated by summing these force terms in continuous space. Although force-based models can describe pedestrian behaviors in detail, they are computationally expensive. Due to the existence of under-damped solutions in second-order ordinary differential equations, sometimes force-based models can produce unrealistic inertia effects such as oscillatory behavior in pedestrian trajectories~\cite{Chraibi_PRE2010, Koster_PRE2013, Kretz_PhysicaA2015}.

\begin{table}
	\caption{Classification of pedestrian models.}
	\label{Model_classification}
	\begin{tabular}{|c|c|}
		\hline
		Macroscopic Models		& Hughes models~\cite{Huang_TrB2009, Hughes_TrB2002}\\
		\hline
		Cellular Automata		& Gipps-Marksj\"{o} model~\cite{Gipps_MCS1985}\\ 
		~						& Blue-Adler model~\cite{Blue_TrB2001}\\
		~						& Floor field model~\cite{Burstedde_PhysicaA2001, Kirchner_PhysicaA2002}\\
		\hline
		Velocity-based Models	& Contractile particle model~\cite{Baglietto_PRE2011}\\ 
		~						& Gradient navigation model~\cite{Dietrich_PRE2014}\\
		\hline
		Force-based Models		& Magnetic force model~\cite{Okazaki_1979}\\
		~						& Social force models~\cite{Helbing_RMP2001, Helbing_PRE1995, Johansson_ACS2007}\\
		~						& Centrifugal force models~\cite{Chraibi_PRE2010, Yu_PRE2005}\\
		\hline
	\end{tabular}
\end{table}

\begin{table}
	\caption{Advantages and limitations of different modeling approaches.}
	\label{compare_models}
	\begin{tabular}{|c|c|c|}
		\hline
		Modeling		& Advantages & Limitations \\
		Approaches		& & \\		
		\hline \hline
		Macroscopic		& computationally cheap &  aggregated descriptions\\
		Models			& & \\		
		\hline
		Cellular		& simple& limited accuracy \\
		Automata		& efficient computing & \\		
		\hline
		Velocity-based	& no inertia effect &  lack of physical interactions\\
		Models			& & \\		
		\hline
		Force-based		& detailed presentations & computationally expensive \\
		Models			&  & inertia effect \\
		\hline
	\end{tabular}
\end{table}

Table~\ref{Model_classification} classifies pedestrian models discussed in this section and Table~\ref{compare_models} summarizes advantages and limitations of different modeling approaches. Interested readers can find extensive reviews of pedestrian flow models presented by Bellomo and Dogbe~\cite{Bellomo_SIAM2011}, Duives~\textit{et al.}~\cite{Duives_TrC2013}, Papadimitriou~\textit{et al.}~\cite{Papadimitriou_TrF2009}, and Templeton~\textit{et al.}~\cite{Templeton_RGP2015}. Bellomo and Dogbe~\cite{Bellomo_SIAM2011} summarized different mathematical models of crowd behaviors and discussed issues on representation scales of the models. They pointed out challenges on developing macroscopic descriptions of collective phenomena based on microscopic models of individual behaviors. Duives~\textit{et al.}~\cite{Duives_TrC2013} compared crowd simulation models and assessed the models' capability for predicting self-organized phenomena in various geometries of indoor space. Papadimitriou~\textit{et al.}~\cite{Papadimitriou_TrF2009} reviewed pedestrian movement models and crossing behavior studies in urban environments. They emphasized the importance of integrating pedestrian movement models with decision-making models such as crossing behavior models. Templeton~\textit{et al.}~\cite{Templeton_RGP2015} examined the underlying assumptions of various crowd behavior models and reported that the majority of crowd behavior models fails to replicate realistic large-scale collective behaviors that have been observed in the previous empirical studies. 

This dissertation has investigated collective patterns of pedestrian flow emerging from interactions between pedestrians and attractions. In order to incorporate behavioral aspects of individuals, the microscopic modeling approach was employed. Among various microscopic models, the social force models have been widely applied, for instance, in agent-based modeling~\cite{Helbing_Nature2000, Oliveira_PRX2016, Parisi_PhysicaA2009} and multiscale modeling~\cite{Corbetta_Dissertation2016, Hoogendoorn_TrC2015, Hoogendoorn_PhysicaA2014}, due to their simplicity. The repulsive force terms in the models are given as natural exponential functions that are easy to integrate and differentiate. The social force models have produced promising results in that the model embodies behavioral elements and physical forces. The original model and its variants have successfully demonstrated various interesting phenomena such as lane formation~\cite{Helbing_PRE1995}, bottleneck oscillation~\cite{Helbing_PRE1995}, and turbulent movement~\cite{Yu_PRE2007}. Accordingly, the social force model was selected for this dissertation. 

\subsubsection{Gaps in the existing social force models}
Different specifications of repulsive force terms have been proposed for the social force models, such as circular specification (CS)~\cite{Helbing_RMP2001}, elliptical specification (ES-1)~\cite{Helbing_PRE1995}, an improved version of the ES-1 (ES-2) introduced by Johansson~\textit{et al.}~\cite{Johansson_ACS2007}. Furthermore, improved numerical implementations of the models have been suggested in order to suppress oscillatory behavior in pedestrian trajectories~\cite{Koster_PRE2013, Kretz_PhysicaA2015}. In addition to the repulsive interactions, the effect of attractive interactions also has been investigated, first introduced by Helbing and Moln\'{a}r~\cite{Helbing_PRE1995}. For example, Xu and Duh~\cite{Xu_IEEE2010} simulated bonding effects of pedestrian groups and evaluated their impacts on pedestrian flows. In another study, Moussa\"{i}d \textit{et al.}~\cite{Moussaid_PLOS2010} analyzed attractive interactions among pedestrian group members and their spatial patterns. However, little attention has been paid to the attractive interactions between pedestrians and attractions.

Based on the review of existing social force models, it is identified that relatively little is known about the effect of attractive interactions between pedestrians and attractions. Therefore, this dissertation aims at proposing numerical simulation models to study the attractive interactions between pedestrians and attractions.

\section{The Concept of Phases}
\label{sec:Phases}
%
Phase is one of the fundamental concepts in thermodynamics and statistical physics~\cite{Schroeder_2000, Sethna_2006}. In these areas, phases indicate different forms of a material or a system of particles. For instance, $\mathrm{H}_2\mathrm{O}$ is in ice, liquid water, or vapor phases depending on temperature and pressure. If a material is in a certain phase, it means that the physical properties of the material are homogeneous even if there are small changes in external conditions. For instance, the specific heat capacity of water is virtually constant if it is in a phase such as water, ice, or vapor. A phase transition can be defined when one can observe a qualitative change from one phase to another by controlling an independent variable. As an example, boiling water shows a phase transition from liquid water to vapor as temperature increases. A phase diagram depicts different phases in a graph as a function of control variables.

The term ``phase'' often indicates different jam patterns in freeway traffic studies. In two-phase traffic theory~\cite{Nagai_PhysicaA2005, Treiber_TrB2010, Wagner_EPJB2008}, traffic flow patterns can be classified into either free flow phase or jammed phase. In the free flow phase, the vehicle density is not high and vehicles can keep large enough distance among them. Although interactions among vehicles exist, the interactions do not significantly reduce the speed of freeway traffic. In the jammed phase, on the other hand, interactions among vehicles come into effect, so movements of a vehicle can be restricted by other vehicles. Consequently, the vehicle speed is decreasing as the vehicle density is growing. Speed drop and stop-and-go patterns can be observed when the freeway traffic turns into the congested phase. According to three-phase traffic theory~\cite{Kerner_2004, Kerner_PRL1997}, the traffic flow patterns are categorized into three phases: free flow, synchronized flow, and wide moving jam phases. In the synchronized flow phase, average speed becomes significantly lower than in the free flow phase, but stopping behavior is not notable yet. The speed drop is often observed at a fixed location and usually does not move upstream. In the wide moving jam phase, however, a congested area propagates upstream and it can move through an area in any state of free or synchronized free phases. In this case, one can observe a coexisting of wide moving jam and either free or synchronized free phases. Stop-and-go patterns can be observed in that vehicles have to stop when they approach the congested area and then speed up when they leave the areas. Furthermore, various jam patterns can be further categorized in terms of multi-phases according to spatio-temporal patterns of the congested area~\cite{Helbing_PRL1999, Schonhof_TrSci2007}. 

In pedestrian flow studies, growing body of literature has mainly focused on jamming transitions, thus the concept of phases is frequently utilized similar to the case of vehicular traffic studies. For example, Ezaki~\textit{et al.}~\cite{Ezaki_PRE2012} and Suzuno~\textit{et al.}~\cite{Suzuno_PRE2013} summarized their study results by presenting jammed phase. Nowak and Schadschneider~\cite{Nowak_PRE2012} studied lane formation in bidirectional flow and suggested four phases including free flow, disorder, lanes, and gridlock phases. Although moving between places is not the only purpose of pedestrian walking, many of pedestrian flow studies have only focused on pedestrian jams in line with pedestrian mobility. Consequently, there remains a need for further research to quantify various patterns or phases in pedestrian motions.

In self-propelled particle studies, the concept of phases has been applied for representing various patterns of collective motions. For instance, D’Orsogna~\textit{et al.}~\cite{DOrsogna_PRL2006} summarized the collective patterns of self-propelled particles based on configurational patterns, such as clumps, ring clumping, and rings. Similar to self-propelled particles, pedestrians can freely move in a two-dimensional space, and their motion shows different patterns not only when they travel but also when they form groups and gather around near an attraction. Therefore, the concept of phases can be applied to explain various patterns of pedestrian motions. In this dissertation, macroscopic measures are introduced to characterize different phases of collective pedestrian behaviors.



\chapter{Numerical Simulation Models }
\label{sec:Models}
In this chapter, I explain numerical simulation models developed in this dissertation. First, I present a numerical simulation model of attractive interactions between pedestrians and attractions. Analogously to self-propelled particle models~\cite{DOrsogna_PRL2006, Mogilner_MathBio2003}, the attractive interactions between pedestrians and attractions are expressed as a superposition of attractive and repulsive force terms. For simplicity, all the pedestrians are subject to the same attractive interactions. The attractive force term makes the pedestrians approach the attractions. The repulsive force term keeps the pedestrians a certain distance from the attractions. Like other social force terms, the attractive force and repulsive force terms in the attractive interaction model are given in an exponential form with the strength and range parameters. Development of the model is described in Section~\ref{sec:model_1}.

Second, I show a joining behavior model which is inspired by the stimulus crowd effect studied by Milgram~\textit{et al.}~\cite{Milgram_JPSP1969} and Gallup~\textit{et al.}~\cite{Gallup_PNAS2012}. Their studies reported that more passersby adapted the behavior of stimulus group as the group size grows. The group members were asked to look up an object on a busy city street for a short time, raising the awareness of the object. Based on their findings, it was assumed that the probability of joining an attraction increases according to the number of pedestrians attending the attraction. If an individual joins an attraction, then he or she stays there for a certain length of time. After the individual leaves the attraction, he or she is not going to visit there anymore. Pedestrian motions were numerically described with the social force model and the joining behavior model. Formulation of the model is presented in Section~\ref{sec:model_2}.

Third, I describe the setup of numerical simulations on pedestrian flow near an attendee cluster. The pedestrians were categorized into two types: attracted pedestrians and passersby. The attracted pedestrians decided to join an attraction according to the joining behavior model presented in Section~\ref{sec:model_2}. The passersby are the pedestrians who are not interested in the attraction, thus they do not visit the attraction. It was assumed that they aimed at smoothly bypassing obstacles while walking towards their destination. The concept of attainable speed was implemented by which each pedestrian adjusted his desired walking speed depending on available space in front of him. Introducing attainable speed of pedestrian motion prevented excessive overlaps among pedestrians. Consequently, it provided a better representation of pedestrian stopping behavior. Details of the numerical simulation model are explained in Section~\ref{sec:model_3}.

Numerical simulations have been performed for a straight corridor of length $L$ along $x$-axis and width $W$ along $y$-axis. It has been assumed that a straight corridor is a basic element of pedestrian facilities, and the pedestrian facilities can be modeled as a combination of such straight corridors. In this dissertation, the corridor length $L$ is considerably larger than the corridor width $W$ in order to make sure that all the pedestrians can see attractions on lower or upper boundaries of the corridor. The study of pedestrian jams induced by an attraction can be performed by setting $W$ small. A long straight corridor provides enough space for a pedestrian queue formed upstream of an attraction. For the straight corridor, bidirectional pedestrian flow was mainly considered. That is, one half of the population is walking towards the right boundary of the corridor from the left, and the opposite direction for the other half. Pedestrians were represented as circles with radius $r_i$. They moved with desired speed $v_d=1.2$ m/s and with relaxation time $\tau=0.5$ s, and their speed was limited to $v_{\rm max}=2.0$~m/s. Here, the desired speed is defined as a speed at which a pedestrian would like to walk if one's walking is not hindered by other pedestrians. The relaxation time controls how fast the pedestrian adapts current speed to desired speed. Pedestrian trajectories were updated with the social force model in Section~\ref{sec:SFM} for each simulation time step $\Delta t = 0.05$~s. Relevant equations for updating pedestrian motions are presented as 
\begin{eqnarray}\label{eq:SFM_general}
	\frac{d\vec{v}_i(t)}{dt} = \vec{f}_{i,d}+\sum_{j\neq i}^{ }{\vec{f}_{ij}}+\sum_{B}^{ }{\vec{f}_{iB}},
\end{eqnarray}
with 
\begin{equation}\label{eq:SFM_terms}
\begin{split}
	\vec{f}_{i,d}	&= \frac{v_d\vec{e}_{i}-\vec{v}_{i}(t)}{\tau},\\
	\vec{f}_{ij}	&= C_p\exp\left(-\frac{b_{ij}}{l_p}\right)\frac{\|\vec{d}_{ij}\|+\|\vec{d}_{ij}-\vec{y}_{ij}\|}{4b_{ij}}\left(\frac{\vec{d}_{ij}}{\|\vec{d}_{ij}\|}+\frac{\vec{d}_{ij}-\vec{y}_{ij}}{\|\vec{d}_{ij}-\vec{y}_{ij}\|}\right) + \vec{g}_{ij},\\
	b_{ij}		&= \frac{1}{2}\sqrt{(\|\vec{d}_{ij}\| + \|\vec{d}_{ij}-\vec{y}_{ij}\|)^{2}-\|\vec{y}_{ij}\|^{2}},\\
	\vec{y}_{ij}	&= (\vec{v}_{j}-\vec{v}_{i}) \Delta t_s,\\
	\vec{f}_{iB}	&= C_b\exp\left(-\frac{d_{iB}}{l_{b}}\right)\vec{e}_{iB}.
\end{split}
\end{equation}

Following previous studies~\cite{Xu_IEEE2010, Zanlungo_EPL2011, Zanlungo_PRE2014}, the first-order Euler method was employed for a numerical integration of Equation~(\ref{eq:SFM_general}). The numerical integration of Equation~(\ref{eq:SFM_general}) was discretized as  
\begin{equation}\label{eq:Euler_method}
	\begin{split}
	\vec{v}_i(t + \Delta t) &= \vec{v}_i(t) + \vec{a}_i(t)\Delta t, \\
	\vec{x}_i(t + \Delta t) &= \vec{x}_i(t) + \vec{v}_i(t + \Delta t)\Delta t.
	\end{split}	
\end{equation}
Here, $\vec{a}_i(t)$ is the acceleration of pedestrian $i$ at time $t$ and the velocity of pedestrian $i$ at time $t$ is given as $\vec{v}_i(t)$. The position of pedestrian $i$ at time $t$ is denoted by $\vec{x}_i(t)$. 

Specific setups of each study are further explained in the following sections. 

\section{Study~\Rmnum{1}: Attractive Force Model}
\label{sec:model_1}
In Study~\Rmnum{1}, the collective effects of attractive interactions between pedestrians and attractions were numerically studied by devising an attractive force term $\vec{f}_{iA}$. The attractive force term $\vec{f}_{iA}$ indicates pedestrian movements toward attractive stimuli, for instance, shop displays and museum exhibits. The attractive force toward attractions was modeled similarly to the interaction between pedestrians and the wall, but in terms of both attractive and repulsive interactions. The repulsive effect of attractions is necessary so that pedestrians can keep a certain distance from attractions. 

Mogilner~\textit{et al.}~\cite{Mogilner_MathBio2003} suggested that attractive interactions among individuals can be modeled by adding repulsion and attraction terms, i.e., 
\begin{eqnarray}\label{eq:MogilnerModel}
	F = C_r \exp \left( -\frac{d_{ij}}{l_r} \right) - C_a\exp \left( -\frac{d_{ij}}{l_a} \right),
\end{eqnarray}
where $F$ is the magnitude of the attractive force between a pair of individuals. The distance between two individuals $i$ and $j$ are represented by $d_{ij}$. Repulsion and attraction strength parameters are indicated by $C_r$ and $C_a$, respectively. Range parameters are denoted by $l_r$ for repulsion and $l_a$ for attraction. They also noted that short-range strong repulsive and long-range weak attractive interactions are needed to make the individuals keep a certain distance between them while forming a stable group. Mogilner~\textit{et al.}~\cite{Mogilner_MathBio2003} and D'Orsogna~\textit{et al.}~\cite{DOrsogna_PRL2006} demonstrated that the pairwise attractive and repulsive forces in Equation~(\ref{eq:MogilnerModel}) can reproduce various formations of swarm particles. 

Analogously to self-propelled particle models~\cite{DOrsogna_PRL2006, Mogilner_MathBio2003}, the strength and range of attractive force were modeled as
\begin{eqnarray}\label{eq:AR}
	\vec{f}_{iA} = \left[C_r\exp\left(\frac{r_i-d_{iA}}{l_r}\right)-C_a\exp\left(\frac{r_i-d_{iA}}{l_a}\right)\right]\vec{e}_{iA}.
\end{eqnarray}
The distance between pedestrian $i$ and attraction $A$ is indicated by $d_{iA}$, and $\vec{e}_{iA}$ is the unit vector pointing from attraction $A$ to pedestrian $i$. Here, $C_a$ and $l_a$ denote the strength and the range of attractive interaction toward attractions, respectively. Similarly, $C_r$ and $l_r$ denote the strength and the range of repulsive interaction from attractions. As in Reference~\cite{Mogilner_MathBio2003}, the strength and range parameters were set to be $C_r > C_a$ and $l_a > l_r$. By these conditions, the attractions attract distant pedestrians but not too close to the attractions. Pedestrian radius $r_i$ is incorporated with the model to consider the effect of pedestrian size on effective distance between pedestrian $i$ and attraction $A$. 


The presented model in Equation~(\ref{eq:AR}) was coupled with the social force model presented in Equation~(\ref{eq:SFM_general}). Accordingly, the social force model was given as 
\begin{eqnarray}\label{eq:SFM_study1}
	\frac{d\vec{v}_i(t)}{dt} = \vec{f}_{i,d}+\sum_{j\neq i}^{ }{\vec{f}_{ij}}+\sum_{B}^{ }{\vec{f}_{iB}}+\sum_{A}^{ }{\vec{f}_{iA}}.
\end{eqnarray}
The interpersonal elastic force $\vec g_{ij}$ in Equation~(\ref{eq:SFM_terms}) was given as 
\begin{eqnarray}
	\vec{g}_{ij} = h(r_{ij}-d_{ij}) \left\{k_n \vec{e}_{ij}+k_t [(\vec{v}_j-\vec v_i)\cdot \vec{t}_{ij}]\vec{t}_{ij}\right\}.
\end{eqnarray}
Other force terms in the implementation were same as in Equation~(\ref{eq:SFM_terms}). 

In numerical simulations, the corridor dimensions were set as $L=25$~m and $W=4$~m, with periodic boundary condition in the direction of $x$-axis. According to the periodic boundary condition, pedestrians walk into the corridor through one side when they have left through the opposite side. Periodic boundary condition was used in order to keep the number of pedestrians constant over the corridor during the numerical simulations. Five attractions were placed for every 5~m on both upper and lower walls of the corridor. To consider the dimension of attractions, each attraction was modeled as three point masses: one point at the center, the other two points at the distance of 0.5 m from the center. The parameters of the repulsive force terms were given based on previous works: $C_p = 3$, $l_p = 0.2$, $\Delta t_{s} = 0.5$, $k_n = 25$, $k_t = 12.5$, $C_b = 10$, and $l_b = 0.2$~\cite{Helbing_TrSci2005, Helbing_PRE1995, Johansson_ACS2007, Moussaid_PNAS2011}. Here, the values of $C_r$ and $l_r$ were set to be the same as $C_b$ and $l_b$. In Study~\Rmnum{1}, short-range strong repulsive and long-range weak attractive interactions were considered, thus the values of $C_a$ and $l_a$ were chosen such that $l_a > l_r$ and $C_a < C_r$. Accordingly, the value of $l_a$ were set to be $1$. The effect of attractive interaction toward attractions can be studied by controlling the ratio of $C_a$ to $C_r$, defining the relative attraction strength $C = C_a/C_r$ with $C_r = 10$. The pedestrian density $\rho=N/A$ was also controlled for the corridor area $A=100$~m$^2$. At the initial time $t=0$, pedestrian positions were randomly generated with uniform distribution over the whole corridor without overlapping.
	
\section{Study~\Rmnum{2}: Joining Behavior Model}
\label{sec:model_2}
Study~\Rmnum{1} examined attractive interactions between pedestrians and attractions by appending the attractive force toward the attractions as shown in Equation~(\ref{eq:AR}). Although the extended social force model can produce various collective patterns of pedestrian movements, Study~\Rmnum{1} did not explicitly take into account selective attention. The selective attention is a widely recognized behavioral mechanism by which people can focus on tempting stimuli and disregard uninteresting ones~\cite{Goldstein_2007, Wickens_1999}. 

Furthermore, it has been widely believed that individual choice behavior can be influenced by social influence from other individuals. For instance, previous studies on stimulus crowd effects reported that a passerby is more likely to shift his attention towards the crowd as its size grows~\cite{Gallup_PNAS2012, Milgram_JPSP1969}. This belief is generally accepted in the marketing area, which can be interpreted that having more visitors in a store can attract more passersby to the store~\cite{Bearden_JCR1989, Childers_JCR1992}. It was also suggested that the sensitivity to others' choice is different for different places, time-of-day, and visitors' motivation~\cite{Gallup_PNAS2012, Kaltcheva_JMkt2006}. In addition, the social influence is also relevant to understand individual choice behavior in emergency evacuations. Helbing~\textit{et al.}~\cite{Helbing_Nature2000} suggested that pedestrians in evacuation scenarios were likely to follow the movement of the majority group due to the lack of information. Kinateder~\textit{et al.}~\cite{Kinateder_TrF2014} discovered that evacuees tended to follow the routes taken by others in virtual reality (VR) experiments of tunnel fires. Based on laboratory experiments of simulated evacuations, Haghani and Sarvi~\cite{Haghani_PhysicaA2017} confirmed that evacuees preferred an exit having more crowds to other exits when the evacuees were unaware of situations of every exit. 

Based on the idea of selective attention~\cite{Goldstein_2007, Wickens_1999} and social influence~\cite{Bearden_JCR1989, Childers_JCR1992, Gallup_PNAS2012, Kaltcheva_JMkt2006, Milgram_JPSP1969}, it was assumed that an individual decided whether he or she was going to visit an attraction based on the number of pedestrians attending the attraction. This is also known as a preferential attachment which can be observed in scientific paper citation~\cite{Price_JASIS1976}, network growth~\cite{Newman_SIAM2003, Sumpter_2010}, and animal group size dynamics~\cite{Nicolis_PRL2013, Sumpter_2010}. According to Gallup~\textit{et al.}~\cite{Gallup_PNAS2012}, the preferential attachment model can be written as
\begin{eqnarray}\label{eq:preferential}
	P(N_i) = \frac{N_{i}^k}{N_{i}^k+N_{j}^k},
\end{eqnarray}
where $P(N_i)$ is the probability of attachment. The number of pedestrians in groups $i$ and $j$ are denoted by $N_i$ and $N_j$, respectively. The exponent $k$ controls the probability function shape. For $k = 1$, the probability function increases as the group size grows and then saturates when the group size is large enough. If $k \geq 2$, the probability function sharply increases before it reaches a particular value, showing a quorum response~\cite{Sumpter_2010}. Similar to the studies of Milgram~\textit{et al.}~\cite{Milgram_JPSP1969} and Gallup~\textit{et al.}~\cite{Gallup_PNAS2012}, the exponent $k$ was set to be 1. For the case of joining behavior, the subscript $i$ can be replaced by $a$ indicating pedestrians already joined an attraction. Likewise, the subscript $j$ can be replaced by $0$ denoting pedestrians who are not stopping by the attraction. In other words, $N_{a}$ and $N_{0}$ are the number of pedestrians who have already joined and for the pedestrians who have not joined the attraction yet, respectively. In addition, $N_a$ was multiplied by social influence parameter $s$ in order to reflect the sensitivity to others' choice. 
\begin{eqnarray}
	P_a = \frac{sN_a}{N_0+sN_a},
\end{eqnarray}
where the social influence parameter $s > 0$ can be also understood as pedestrians' awareness of the attraction. The above equation shows that the choice probability increases when $s$ or $N_a$ grows. However, the choice probability cannot be estimated if there is nobody within the range of perception. In order to avoid such indeterminate case, $K_0$ and $K_a$ were introduced as baseline values of $N_{a}$ and $N_{0}$, respectively, i.e., 
\begin{eqnarray}\label{eq:P_a}
	P_{a} = \frac{s(N_{a}+K_a)}{({N_{0}+K_0})+s({N_{a}+K_a})}.
\end{eqnarray}
According to previous studies~\cite{Gallup_PNAS2012, Kaltcheva_JMkt2006, Milgram_JPSP1969}, it was assumed that the strength of social influence can be different for different situations and can be controlled in Equation~(\ref{eq:P_a}). When there is nobody near the attraction (i.e., $N_a$ and $N_0$ are 0), the joining probability becomes $P_a = sK_a/(K_0+sK_a)$. In this case, the joining probability depends on the values of $K_0$ and $K_a$, meaning that an individual is more likely join the attraction as the value of $K_a$ increases. 

When an individual can see an attraction in the perception range $R_i = 10$~m, the individual evaluates the joining probability in Equation~(\ref{eq:P_a}). If the individual decides to join the attraction, then he or she shifts desired direction vector $\vec e_i$ toward the attraction. Once an individual has joined an attraction, he or she will stay near the attraction for an exponentially distributed time with an average duration of visiting an attraction $t_{d}$~\cite{Gallup_PNAS2012, Helbing_PRE1995, Kwak_PLOS2015}. After a lapse of $t_d$, the individual leaves the attraction and continues walking towards initial destination, not visiting the attraction again. While $t_d$ was observed on the order of a second in References~\cite{Gallup_PNAS2012, Milgram_JPSP1969}, $t_d$ was a control variable in Study~\Rmnum{2} ranging up to 300~s in order to consider different types of attractions in terms of $t_d$.

Pedestrian motions were updated with the social force model presented in Equation~(\ref{eq:SFM_general}). Comparing to the setup of Study~\Rmnum{1}, a simpler form of interpersonal elastic force $\vec g_{ij}$ in Equation~(\ref{eq:SFM_terms}) was given as in Reference~\cite{Moussaid_PNAS2011}, i.e., 
\begin{eqnarray}
	\vec{g}_{ij} = k_n h(r_{ij}-d_{ij}) \vec{e}_{ij}.
\end{eqnarray}
Other force terms in the implementation are same as in Equation~(\ref{eq:SFM_terms}).

In numerical simulations, the corridor dimensions were set as $L=30$~m and $W=6$~m, with periodic boundary condition in the direction of $x$-axis. An attraction was placed at the center of the lower wall, i.e., at the distance of 15~m from the left boundary of the corridor. The number of pedestrians in the corridor, $N$, was given as 100. As in Study~\Rmnum{1}, pedestrian positions were randomly generated with a uniform distribution over the whole corridor without overlapping. The parameters of the repulsive force terms were given based on previous works: $C_p = 3$, $l_p = 0.2$, $\Delta t_{s} = 0.5$, $k = 62.5$, $C_b = 10$, and $l_b = 0.2$~\cite{Helbing_PRE1995, Johansson_ACS2007, Kwak_PRE2013, Moussaid_PNAS2011}. For simplicity, $K_a$ and $K_0$ were set to be 1, meaning that both options are equally attractive when the individual would see nobody within his perception range. An individual was counted as an attending pedestrian if his efficiency of motion $E_{i} = (\vec{v}_{i} \cdot \vec{e}_{i})/{v_d}$ was lower than 0.05 within a range of 3~m from the center of the attraction after he decided to join there. The efficiency of motion indicates how much the driving force contributes to the progress of pedestrian $i$ towards his destination with a range from 0 to 1~\cite{Helbing_PRL2000, Kwak_PRE2013}. Here, $E_{i} = 1$ implies that the individual is walking towards his destination with the desired velocity while lower $E_{i}$ indicates that an individual is distracted from his initial destination because of the attraction. 

\section{Study~\Rmnum{3}: Modeling Pedestrian Flow near an Attendee Cluster}
\label{sec:model_3}
In Study~\Rmnum{3}, the corridor dimensions were set as $L=60$~m and $W=4$~m, with an open boundary condition in the direction of $x$-axis. An attraction was placed at the center of the lower wall, i.e., at the distance of 30~m from the left boundary of the corridor. An open boundary condition was employed in order to continuously supply passersby to the corridor. By doing so, pedestrians can enter the corridor regardless of the number of leaving pedestrians. On the other hand, if the number of pedestrians in the corridor is fixed by periodic boundary condition, the number of passersby tends to decrease as the number of attracted pedestrians increases. 

The number of pedestrians in the corridor was associated with the pedestrian influx $Q$, i.e., the arrival rate of pedestrians entering the corridor. The unit of $Q$ is indicated by P/s, which stands for pedestrians per second. Based on previous studies~\cite{Luttinen_TRR1999, May_1990}, the pedestrian inter-arrival time was assumed to follow a shifted exponential distribution. That is, pedestrians were entering the corridor independently and their arrival pattern was not influenced by that of others. The minimum headway was set to $0.4$~s between successive pedestrians entering the corridor, which is large enough to prevent overlaps between arriving pedestrians. 

Pedestrians were entering the corridor through the left and the right boundaries in case of bidirectional flow scenario. The corridor boundaries were sliced into inlets with a width of 0.5~m which is slightly larger than pedestrian size (i.e., $2r_i = 0.4$~m), so there are 8 inlets on each boundary. In each inlet, pedestrians were entering the corridor unassociated with entering pedestrians in the neighboring inlets. For each entering pedestrian, the vertical position was randomly generated based on uniform distribution within the entering inlet without overlapping with other nearby pedestrians and boundaries. In effect, pedestrians were inserted at random places on either side.

In addition to bidirectional flow scenario, unidirectional flow was also considered because one can observe different collective patterns of passersby that were not observed in the bidirectional flow. In unidirectional flow scenario, all the pedestrians were entering the corridor through the left boundary and walking towards the right. 

Similar to Study~\Rmnum{2}, pedestrian motions were updated with the social force model in Equation~(\ref{eq:SFM_general}). However, the formulation of desired speed $v_d$ in driving force term $f_{i, d}$ was modified in order to provide a better representation of pedestrian stopping behavior. Inspired by the previous studies~\cite{Chraibi_PRE2015, Parisi_PhysicaA2009}, it was assumed that the desired speed $v_d$ is an attainable speed of pedestrian $i$ depending on the available walking space in front of the pedestrian, 
\begin{eqnarray}\label{eq:desired_speed}
	v_d = \min\{v_0, d_{ij}/T_c\},
\end{eqnarray}
where $v_0$ is a comfortable walking speed and $d_{ij}$ is the distance between pedestrian $i$ and the first pedestrian $j$ encountering with pedestrian $i$ in the course of $\vec{v_i}$. Time-to-collision $T_c$ represents how much time remains for a collision of two pedestrians $i$ and $j$. In line with previous studies~\cite{Karamouzas_PRL2014, Moussaid_PNAS2011}, time-to-collision $T_c$ was estimated by extending current velocities of pedestrians $i$ and $j$, $v_i$ and $v_j$, from their current positions, $x_i$ and $x_j$:
\begin{eqnarray}\label{eq:T_c}
	T_c = \frac{\beta - \sqrt{\beta^2 - \alpha\gamma}}{\alpha},
\end{eqnarray}
where $\alpha = \left\| \vec{v}_i - \vec{v}_j \right\|^2$, $\beta = (\vec{x}_i-\vec{x}_j)\cdot(\vec{v}_i-\vec{v}_j)$, and $\gamma = \left\| \vec{x}_i-\vec{x}_j \right\|^2 - (r_i + r_j)^2$. Note that $T_c$ is valid for $T_c > 0$, meaning that pedestrians $i$ and $j$ are in a course of collision, whereas $T_c < 0$ implies the opposite case. If $T_c = 0$, disks of pedestrians $i$ and $j$ are in contact.

It was assumed that passersby aim at smoothly bypassing obstacles while walking towards their destination. Note that passersby in Study~\Rmnum{3} are the pedestrians who are not interested in the attraction, thus they do not visit the attraction. Analogously to the potential flow in fluid dynamics~\cite{Anderson_2010, Batchelor_2000} and in pedestrian stream model~\cite{Hughes_TrB2002}, the streamline function was employed to steer passersby between boundaries of the corridor. The streamlines represent plausible trajectories of particles smoothly bypassing obstacles, and the partial derivatives of the streamlines can express the flow velocity components in x- and y-directions for a location $z = (x, y)$. For passerby flow moving near an attraction, an attendee cluster can act as an obstacle. It was assumed that passersby set their initial desired walking direction $\vec{e}_{i, 0}$ along the streamlines. As reported in a previous study~\cite{Kwak_PLOS2015}, the shape of an attendee cluster near an attraction can be approximated as a semicircle. By doing so, one can set the streamline function $\psi$ for passerby traffic similar to the case of fluid flow around a circular cylinder in a two dimensional space~\cite{Anderson_2010, Batchelor_2000}:
\begin{eqnarray}\label{eq:streamline}
	\psi = v_0 d_{zA} \sin(\theta_z) \left(1- \frac{r_c}{d_{zA}} \right), %
\end{eqnarray}
where $v_0$ is the comfortable walking speed and $d_{zA}$ is the distance between the center of the semicircle $A$ and location $z = (x, y)$. The angle $\theta_z$ was measured between $y = 0$~m and $\vec{d}_{zA}$. The attendee cluster size at time $t$ is denoted by $r_c = r_c(t)$. To measure $r_c$, the walking area near the attraction was sliced into thin layers with the width of a pedestrian size (i.e., $2r_i = 0.4$~m) in the horizontal direction. From the bottom layer to the top layer, one can count the number of layers consecutively occupied by attendees. The attendee cluster size $r_c$ was then obtained by multiplying the number of consecutive layers by the layer width $0.4$~m. The initial desired walking direction $\vec{e}_{i, 0}$ can be obtained as 
\begin{eqnarray}\label{eq:e_i0}
	\vec{e}_{i, 0} = \left( \frac{\partial \psi}{\partial y}, -\frac{\partial \psi}{\partial x} \right).
\end{eqnarray}
Note that passersby pursue their initial destination, thus their desired walking direction $\vec{e}_{i}$ is identical to $\vec{e}_{i, 0}$ given in Equation~(\ref{eq:e_i0}), i.e, $\vec{e}_{i} = \vec{e}_{i, 0}$.

The repulsive force $\vec{f}_{ij}$ was also modified to simulate the motions of attracted pedestrians near an attraction and passersby who do not visit the attraction. The anisotropic function $\omega_{ij}$ was included in $\vec{f}_{ij}$ in order to represents pedestrian $i$'s directional sensitivity to pedestrian $j$~\cite{Johansson_ACS2007}. The modified repulsive force $\vec{f}_{ij}$ is given as 
\begin{eqnarray}\label{eq:fij_study3}
	\vec{f}_{ij} = \omega_{ij}C_p\exp\left(-\frac{b_{ij}}{l_p}\right)\frac{\|\vec{d}_{ij}\|+\|\vec{d}_{ij}-\vec{y}_{ij}\|}{4b_{ij}}\left(\frac{\vec{d}_{ij}}{\|\vec{d}_{ij}\|}+\frac{\vec{d}_{ij}-\vec{y}_{ij}}{\|\vec{d}_{ij}-\vec{y}_{ij}\|}\right)	
\end{eqnarray}
with 
\begin{eqnarray}\label{eq:anisotropic_study3}
	\omega_{ij}	= \lambda_{ij} + (1-\lambda_{ij})\frac{1+\cos~\phi_{ij}}{2}.
\end{eqnarray}
Here, $0 \leq \lambda_{ij} \leq 1$ is pedestrian $i$'s minimum anisotropic strength against pedestrian $j$. The angle $\phi_{ij}$ was  measured between velocity vector of pedestrian $i$, $\vec{v}_{i}$, and relative location of pedestrian $j$ with respect to pedestrian $i$, $\vec{d}_{ji} = \vec{x}_j - \vec{x}_i$. In Study~\Rmnum{3}, pedestrians were represented as non-elastic solid discs in order to mimic pedestrian stopping behavior. Thus, interpersonal elastic force $\vec g_{ij}$ compression among pedestrians was not modeled.

The parameters of the repulsive force terms were given based on previous works: $C_p = 3$, $l_p = 0.3$, $\Delta t_s = 2.5$, $C_b = 6$, and $l_b = 0.3$~\cite{Helbing_PRE1995, Johansson_ACS2007, Kwak_PRE2013, Kwak_PLOS2015, Zanlungo_PLOS2012}. The minimum anisotropic strength $\lambda_{ij}$ was set to 0.25 for attendees near the attraction and 0.5 for others, yielding that the attendees exert smaller repulsive force on others than passersby do. Consequently, the attendees can stay closer to the attraction while being less disturbed by the passersby. 

As in Study~\Rmnum{2}, an individual evaluates the joining probability when the attraction can be seen by the individual $10$~m ahead. The joining behavior model was implemented as in Study~\Rmnum{2}, but the average duration of visiting an attraction $t_d$ was set to be $30$~s. While the influence of $t_d$ on pedestrian visiting behavior was one of the main interests in Study~\Rmnum{2}, Study~\Rmnum{3} focused on the impact of $Q$ and $s$ on passerby traffic flow near an attraction. Setting up the value of $t_d$ as $30$~s enables one to observe enough number of passersby and attendees near the attraction. If $t_d$ is too small, an attendee cluster might not be observed. On the other hand, if $t_d$ is too large, passerby flow might not exist because large $t_d$ tends to produce a large attendee cluster which can block the corridor.  

\chapter{Results}
\label{sec:Results}
\section{Study~\Rmnum{1}: Collective Dynamics of Attracted Pedestrians}

\begin{figure}[!t]
	\centering
	\subfigure{}
	\includegraphics[width=10cm]{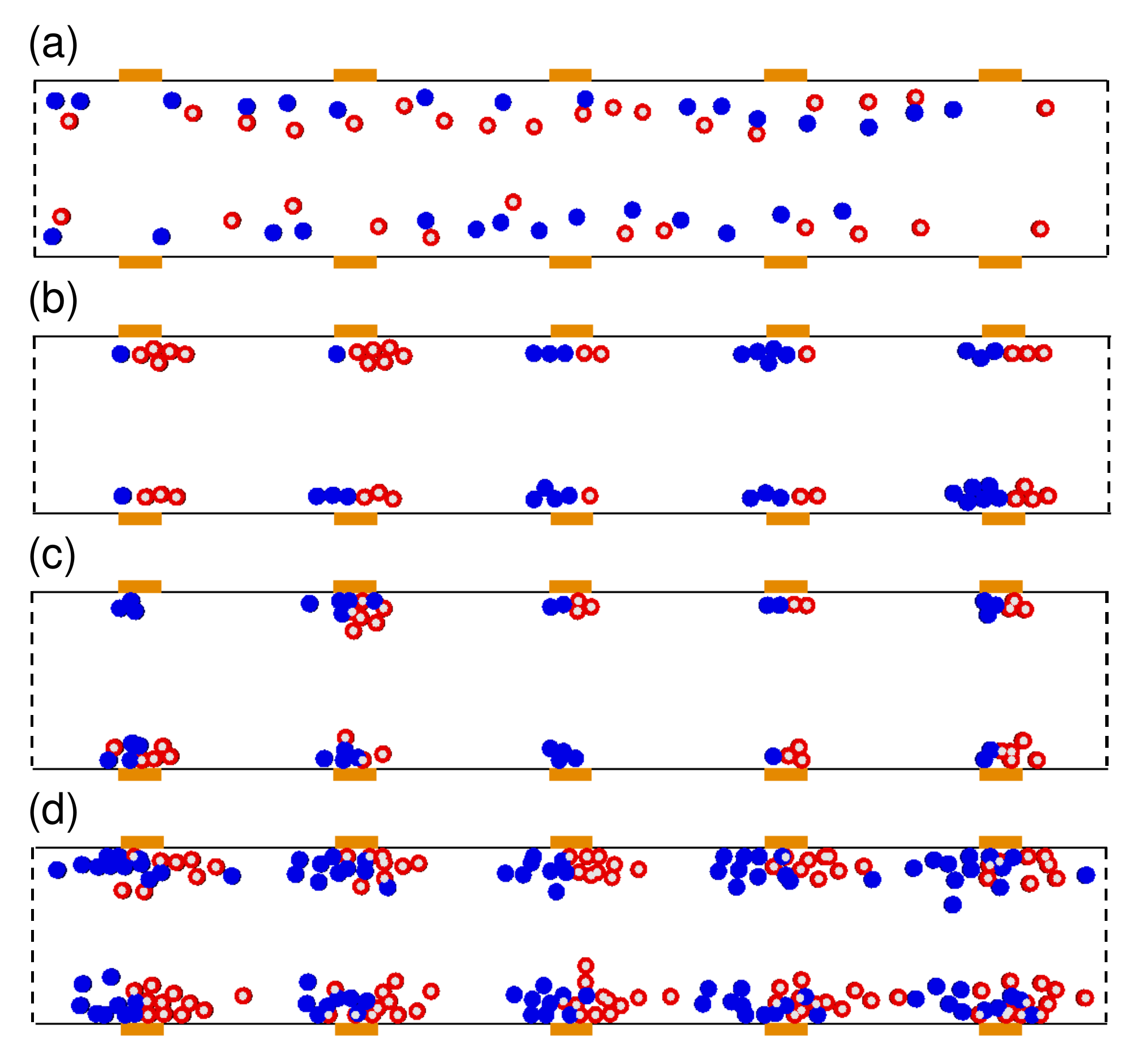}
	\centering
	\subfigure{}
	\includegraphics[width=9cm]{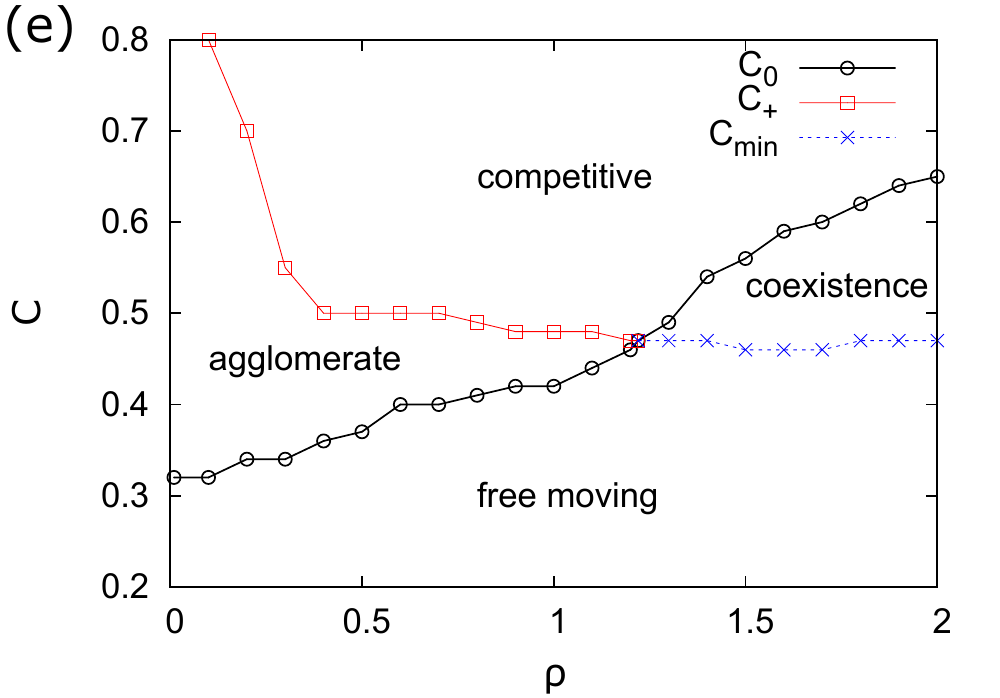}
	\caption{(color online) Snapshots of numerical simulations for various collective patterns: (a) free moving, (b) agglomerate, (c) competitive phase, and (d) coexistence subphase. The attractions, depicted by rectangles, are located on the walls of the corridor with periodic boundary conditions in the horizontal direction. Blue and red circles depict the pedestrians with desired directions to the left and to the right, respectively. (e) Phase diagram summarizing the numerical results. The parameter space of the pedestrian density $\rho$ and the relative attraction strength $C$ is divided into four regions by means of $C_0$ ($\circ$), $C_+$ ($\Box$), and $C_{\rm min}$ ($\times$). Here, $C_0$ shows the critical point at which the efficiency of motion $E$ becomes zero. $C_+$ denotes parameter combination $(\rho, C)$ at which the normalized kinetic energy $K$ begins to grow from 0 while $E = 0$. $C_{\rm min}$ indicates parameter combination $(\rho, C)$ at which $\partial K/\partial C > 0$ while $E > 0$.}
	\label{fig:results_1} 
\end{figure}

To study collective dynamics of attractive interactions between pedestrians and attractions, the attractive force model presented in Section~\ref{sec:model_1} was implemented and examined in Study~\Rmnum{1}. Extensive numerical simulations were performed by controlling relative attraction strength $C$ and pedestrian density $\rho$. 

One can predict four representative patterns depending on $C$ and $\rho$. When $\rho$ is low, three different phases are observed. For small $C$, pedestrians walk to their destinations without being influenced by attractions, which shows a free flow phase (see Figure~\ref{fig:results_1}(a)). For intermediate values of $C$, the attractive interactions lead to an agglomerate phase where pedestrians form stable clusters around attractions, as shown in Figure~\ref{fig:results_1}(b). When $C$ is strong, one can observe a competitive phase characterized by pedestrians rushing into attractions and pushing others; see Figure~\ref{fig:results_1}(c). If $\rho$ is high, the agglomerate phase is not observed anymore. Instead, the free moving and competitive phases coexist only for the intermediate attractive interaction (see Figure~\ref{fig:results_1}(d)). 

These collective patterns of pedestrian movements or phases were characterized by introducing the efficiency of motion $E$ and the normalized kinetic energy $K$~\cite{Helbing_PRL2000} as following:
\begin{eqnarray}\label{eq:E}
	E = \left\langle \frac{1}{N} \sum_{i=1}^{N} \frac{\vec{v}_{i}\cdot\vec{e}_{i}}{v_d} \right\rangle
\end{eqnarray}
and
\begin{eqnarray}\label{eq:K}
	K = \left\langle \frac{1}{N} \sum_{i=1}^{N} \frac{\|\vec{v}_{i}\|^2}{v_d^2} \right\rangle.
\end{eqnarray}
Here $\langle\cdot\rangle$ represents an average over 60 independent simulation runs after reaching the stationary state. The efficiency reflects the contribution of the driving force in the pedestrian motion. If all the pedestrians walk with their desired velocity, the efficiency becomes $1$. On the other hand, the zero efficiency can be obtained if pedestrians do not move in their desired directions and form clusters at attractions. The normalized kinetic energy has the value of $0$ if all the pedestrians do not move, otherwise it has a positive value.

Based on those macroscopic measures, different phases were identified. For a given $\rho$, $E$ decreased according to $C$, meaning that pedestrians are more distracted from their desired velocity due to the larger strength of attractions. Interestingly, $E$ became $0$ at a finite value of $C = C_0(\rho)$, indicating the transition from the free moving phase to the agglomerate phase for low $\rho$ or to the competitive phase for high $\rho$. As $C$ increased, $K$ turned to be $0$ at the same critical point of $C = C_0$, and then gradually increased from $0$ at $C = C_+(\rho)$ with $C_+ \geq C_0$. The boundaries among different phases were characterized in terms of $C_0$ and $C_+$. For low values of $\rho$, the free moving phase for $C < C_0$ was characterized by 
\begin{equation}
	E > 0\ \textrm{and}\ K > 0,
\end{equation}
indicating that kinetic energy of pedestrian motion is mostly used for progressing towards their destinations. The upper boundary of free flow phase was identified by $C = C_0(\rho)$ at which $E$ becomes $0$. For $C_0 < C < C_+$, the agglomerate phase was characterized by 
\begin{equation}
	E = K = 0,
\end{equation}
showing that pedestrians came to a standstill. When $C > C_+$, the competitive phase by 
\begin{equation}
	E = 0\ \textrm{and}\ K > 0,
\end{equation}
reflecting that pedestrians do not walk in their desired directions but still moving near the attractions. For high values of $\rho$, however, $K$ decreased and then increased according to $C$ but without becoming zero. The striking difference from the case with low $\rho$ is the existence of parameter region characterized by both 
\begin{equation}
	E > 0\ \textrm{and}\ \frac{\partial K}{\partial C} > 0.
\end{equation}
This implies that some pedestrians rush into attractions as in the competitive phase (i.e., $\partial K/\partial C > 0$), while other pedestrians move in their desired directions as in the free moving phase (i.e., $E > 0$). The moving pedestrians are also attracted by attractions but cannot stay around them because of interpersonal repulsion effect by other pedestrians closer to attractions. Thus, this parameter region was characterized as a coexistence subphase. The coexistence subphase belongs to the free moving phase in the sense that the coexistence subphase was also characterized by $E > 0$ and $K > 0$. However, one can identify the lower boundary of coexistence subphase by determining $C_{\rm min}$ that minimizes $K$. The upper boundary to the competitive phase was determined by the critical point $C_0$. It was found that $C_0(\rho)$ is an increasing function of $\rho$ because stronger attractions are needed to entice more pedestrians. Different phases characterizing different collective patterns of pedestrians and transitions among them were summarized in the phase diagram of Figure~\ref{fig:results_1}(e). 

The appearance of various phases in Study~\Rmnum{1} can be explained by the interplay between attraction strength and interpersonal repulsion effect. If the pedestrian density is low, one can observe the transition from the free phase to the agglomerate phase and finally to the competitive phase. As $C$ increases, pedestrians tend to be more distracted from their velocity due to the larger strength of attractions, yielding to the appearance of the agglomerate phase. After the agglomerate phase emerges, further increasing $C$ leads to competitive phase. In this phase, attracted pedestrians jostle each other because of the interpersonal repulsion effect. As $C$ increases, their jostling behavior becomes more severe because higher $C$ increases the desire to reaching the attractions, leading to smaller interpersonal distance. Consequently, the interpersonal repulsion effect becomes critical. For high pedestrian density, the agglomerate phase does not appear and the coexistence subphase emerges. The moving pedestrians are also attracted by the attractions but cannot stay around the attractions. This is due to the interpersonal repulsion effect by other pedestrians closer to the attractions. 

\section{Study~\Rmnum{2}: Visiting Behavior}

\begin{figure}[!t]
	\centering
	\subfigure{}
	\includegraphics[width=10cm]{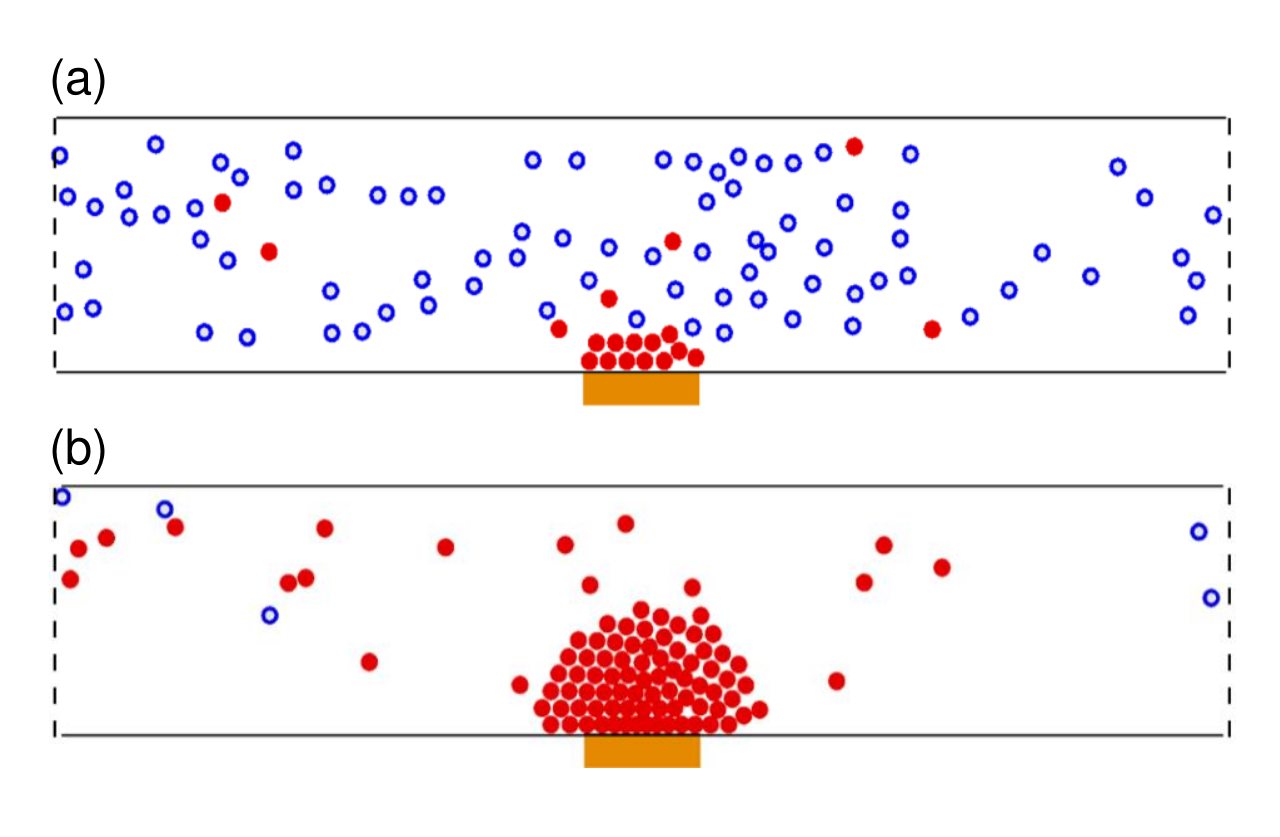}\vspace{-0.5cm}
	\subfigure{}
	\includegraphics[width=9.5cm]{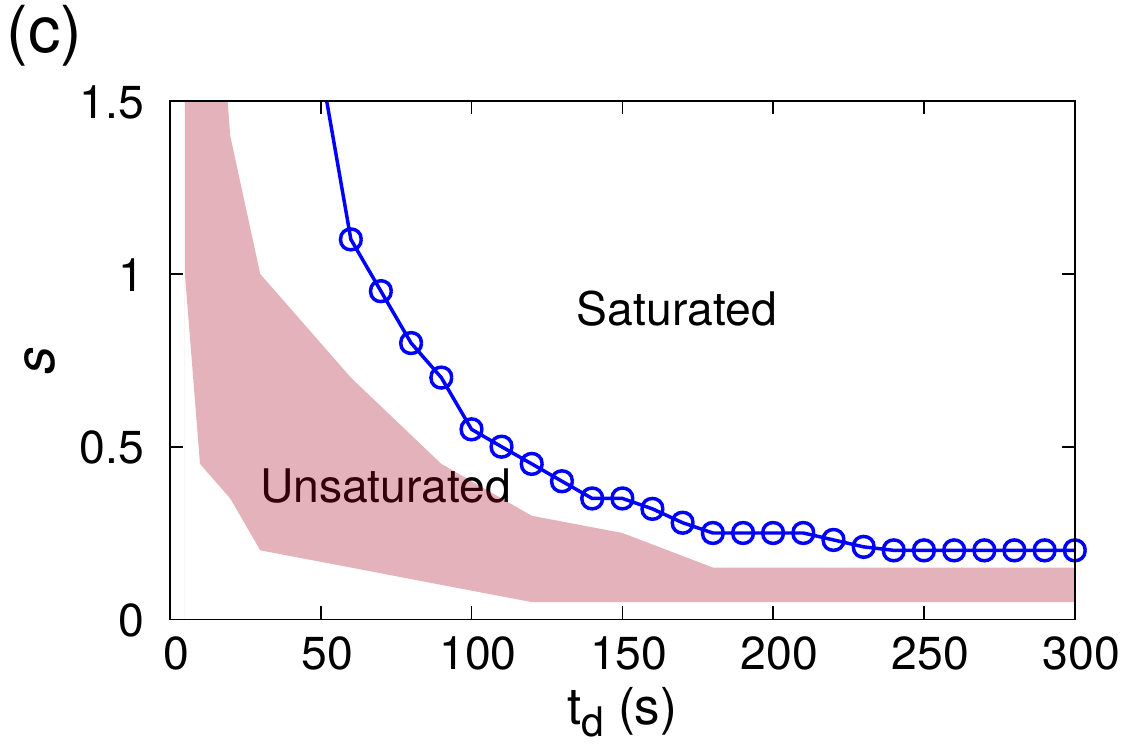}
	\caption{(color online) Snapshots of numerical simulations for different collective patterns: (a) unsaturated phase and (b) saturated phase. The attraction, depicted by an orange rectangle, is located at the center of the lower wall. Filled red and hollow blue circles depict the pedestrians who have and have not visited the attraction, respectively. (c) A feasibility zone of effective improvements and an envelope of the saturated phase. The red shaded area depicts a feasibility zone of the effective improvements characterized in terms of the marginal benefits. The blue solid line with circle symbols represents the envelope of the saturated phase.}
	\label{fig:results_2} 
\end{figure}

In Study~\Rmnum{2}, the pedestrian joining behavior model in Section~\ref{sec:model_2} was implemented and numerically studied in order to understand the collective patterns of pedestrians' visiting behavior. Note that while pedestrians in Study~\Rmnum{1} were subject to the same equation of motion, the joining behavior model in Study~\Rmnum{2} enabled pedestrians to make decisions between walking in the desired direction and stopping by an attraction. The numerical simulation results showed different patterns of pedestrian movements depending on the social influence strength $s$ and the average duration of visiting an attraction $t_{d}$. If $s$ is weak, one can define an unsaturated phase where some pedestrians move towards the attraction while others walk in their desired directions; see Figure~\ref{fig:results_2}(a). For large values of $s$, a saturated phase can be defined, in which every pedestrian near the attraction is joining or has visited the attraction, as shown in Figure~\ref{fig:results_2}(b). These different patterns were characterized by comparing $N_p$ and $N_v$. One can define $N_p$ as the number of pedestrians near the attraction, i.e., within a range of $R_a = 10$~m from the center of the attraction, and $N_v$ as the number of pedestrians near the attraction who have visited the attraction. Here, a pedestrian is considered as ``having visited an attraction'' if one is joining there or one has left the attraction after having stayed there for $t_{d}$. The saturated phase was characterized by 
\begin{equation}
	N_v = N_p,
\end{equation}
indicating that all the pedestrians near the attraction have visited the attraction. On the other hand, the unsaturated phase was characterized by
\begin{equation}
	N_v < N_p,
\end{equation}
where some pedestrians have visited the attraction while others walk in their desired directions.

In addition to describing observed visiting patterns, two marginal benefits of facility improvements were also evaluated to reflect the change in $N_v$. It was assumed that facility improvements can be achieved by either increasing the social influence strength $s$ or the average duration of visiting an attraction $t_{d}$. Accordingly, the marginal benefits were calculated as the first derivate of $N_v$ with respect to $s$ and $t_{d}$, respectively. If the marginal benefits are significant, the corresponding parameter region of $s$ and $t_d$ is denoted by a feasibility zone of the effective improvements. That is, higher increase of $N_v$ is expected with smaller increase of $s$ and $t_d$. Other than the feasibility zone, the impact of changing these variables are insignificant. Figure~\ref{fig:results_2}(c) summarizes the finding in Study~\Rmnum{2}.

As can be seen from Figure~\ref{fig:results_2}(c), the parameter space of $s$ and $t_d$ was divided into two regions according to the value of $N_v$. For each value of $t_{d} > 43$ s, $N_v$ increases as $s$ increases, and then finally reaches $N_p$ at a critical value of $s$, i.e., $s_c$. This indicates the transition from the unsaturated phase to the saturated phase. For the region of $t_{d} < 43$~s, $N_v$ increases as $s$ grows but does not reach to its maximum allowed value even for the large values of $s$. The average duration of visiting an attraction $t_d$ is not long enough, so it fails to obtain sufficient amount of visitors. One can also observe that $s_c$ substantially decreases for $t_d \leq 120$~s and appears to stay around $0.2$ when $t_d$ is larger than $180$~s. It is reasonable to suppose that the departure rate of attending pedestrians is positively associated with the reciprocal of $t_d$ while the arrival rate of joining pedestrians is linked to $s$. Accordingly, one can infer that the impact of the departure rate on $N_v$ is dominated by that of the arrival rate when $t_d$ is large, so the marginal impact of increasing $t_d$ becomes less notable for larger $t_d$.

\section{Study~\Rmnum{3}: Jamming Transitions Induced by an Attraction}
Study~\Rmnum{3} investigated jamming transitions in pedestrian flow interacting with an attraction by means of numerical simulations. Similar to Study~\Rmnum{2}, the simulation model was mainly based on the social force model presented in Chapter~\ref{sec:Models} for pedestrian motions and the probability of joining an attraction presented. In order to provide a better representation of pedestrian stopping behavior, formulations of the desired speed $v_d$ and desired walking direction vector $e_i$ were further improved as presented in Section~\ref{sec:model_3}. By controlling pedestrian influx $Q$ and the social influence parameter $s$, various pedestrian flow patterns were observed in unidirectional and bidirectional flow scenarios. Again, note that the unit of $Q$ is indicated by P/s, which stands for pedestrians per second.

\subsection{Unidirectional Flow}

\begin{figure}[!t]
	\centering
	\subfigure{}
	\includegraphics[width=10cm]{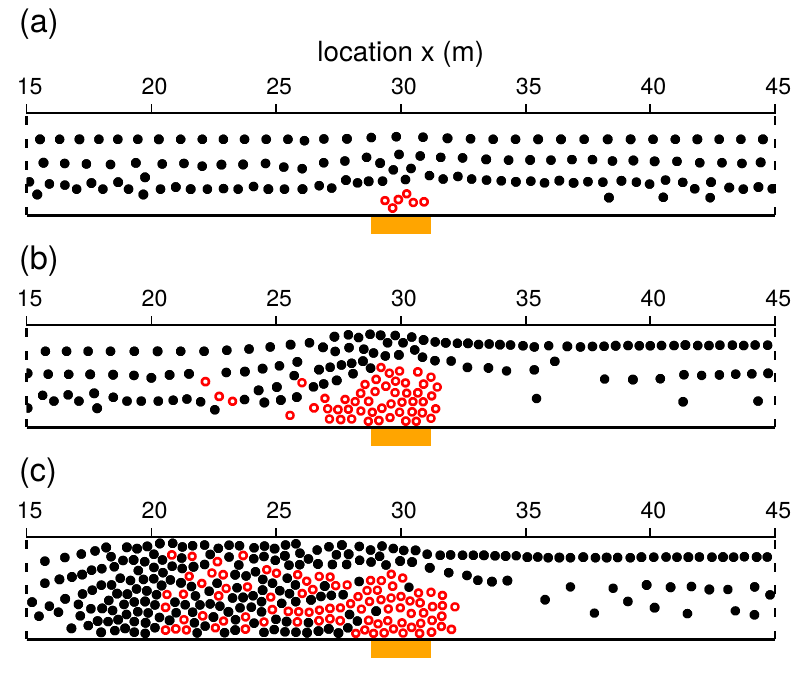}
	\centering
	\subfigure{}
	\includegraphics[width=5.5cm]{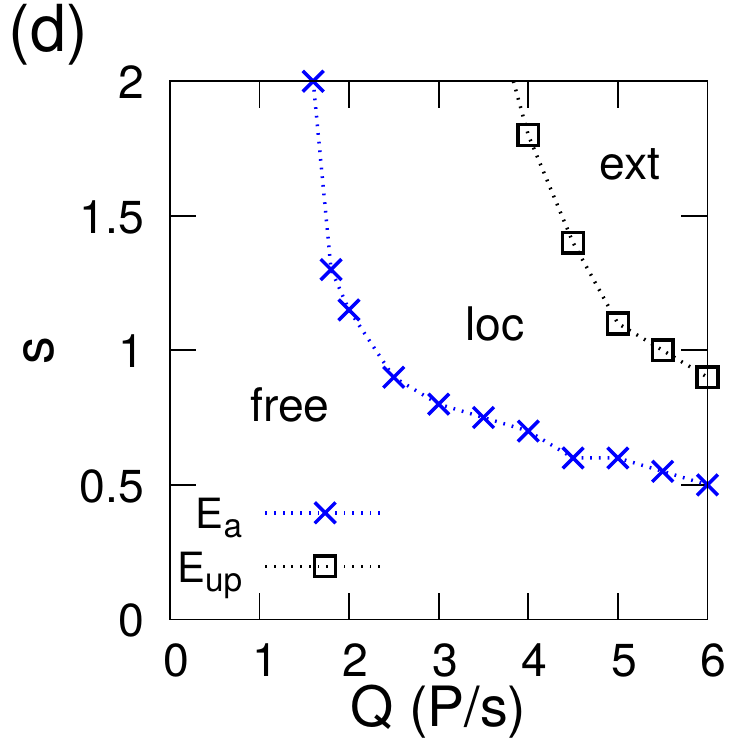}\vspace{-0.3cm}
	\caption{(color online) Representative snapshots of different passerby flow patterns in unidirectional flow: (a) free flow phase, (b) localized jam phase, and (c) extended jam phase. The attraction, depicted by an orange rectangle, is located at the center of the lower wall with open boundary conditions in the horizontal direction. Passersby walking from the left to the right are indicated by filled black circles. Hollow red circles depict pedestrians attracted by the attraction. (d) Phase diagram summarizing the numerical results of unidirectional flow. Here, `free', `loc', and `ext' indicate the free flow phase, the localized jam phase, and the extended jam phase, respectively. Note that P/s stands for pedestrians per second, being the unit of pedestrian flux $Q$.}
	\label{fig:results_3uni} 
\end{figure}

In unidirectional flow, a free flow phase was defined for small $Q$ and $s$, see Figure~\ref{fig:results_3uni}(a). A localized jam phase appeared in the vicinity of the attraction for medium and high $Q$ with the intermediate range of $s$, as can be seen from Figure~\ref{fig:results_3uni}(b). Passersby walked slowly near the attraction because of the reduced walking area, and then they recovered their speed after they walked away from the attraction. One can observe that pedestrians walking away from the attraction tend to form lanes. This is possible because the standard deviation of speed among the walking away pedestrians is not significant after the pedestrians recover their speed. According to the study of Moussa\"{i}d~\textit{et al.}~\cite{Moussaid_PLOS2012}, the formation of pedestrian lanes is stable when pedestrians are walking at nearly the same speed. Once the walking away pedestrians form lanes, the lanes are not likely to collapse. An extended jam phase was observed when both $Q$ and $s$ were large. In the extended jam phase, the pedestrian queue was growing towards the left boundary and the queue was persisting for a long period of time; see Figure~\ref{fig:results_3uni}(c). The attendee cluster did not maintain its semi-circular shape anymore. This seems to be because passersby seized up the attracted pedestrians. Meanwhile, pedestrians in the queue still could slowly walk towards the right side of the corridor as they initially intended. When $Q$ and $s$ were very large, the extended jam phase could end up in a freezing phenomenon with a certain probability, indicating that passersby could not proceed beyond the attraction due to the clogging effect. Some passersby were pushed out towards the attraction by the attracted pedestrians and inevitably they prevented attracted pedestrians from joining the attraction. Consequently, the attracted pedestrians could not approach the boundary of the attendee cluster although they kept their walking direction towards the attraction. Simultaneously, the passersby near the attraction attempted to walk away from the attraction, but they could not because they were blocked by the attracted pedestrians. Eventually, the pedestrian movements near the attraction came to a halt.

Based on observations presented in Figures~\ref{fig:results_3uni}(a) to \ref{fig:results_3uni}(c), different passerby flow patterns were characterized in terms of stationary state average of local efficiency, $E(x)$:
\begin{eqnarray}\label{eq:E_x}
	E(x) = \langle E(x, t) \rangle =\left\langle  \frac{1}{|N_{p}(x, t)|} \sum_{i\in N_{p}(x, t)}{E_i} \right\rangle,
\end{eqnarray}
where $\langle\cdot\rangle$ represents the average obtained from 50 independent simulation runs after reaching the stationary state. The individual efficiency of motion $E_i = (\vec{v}_{i} \cdot \vec{e}_{i, 0})/{v_0}$ can be understood as a normalized speed of pedestrian $i$ in the horizontal direction. The local efficiency $E(x, t)$ indicates how fast passersby in segment $x$ progress towards their destination at time $t$. The set $N_{p}(x, t)$ is the collection of passersby in a $1$~m long segment $x$ at time $t$, and $|N_{p}(x, t)|$ is the cardinality of the set $N_{p}(x, t)$. If $|N_{p}(x, t)| = 0$, $E(x,t)$ was set to be $1$ inferring that a passerby can walk with comfortable speed $v_0$ if the passerby is in the segment $x$ at time $t$. When $E(x) = 1$, the passersby can freely walk without reducing their speed. In contrast, $E(x) = 0$ implies that the passersby have reached a standstill.

A section of $27~\text{m} \leq x \leq 33~\text{m}$ was selected to evaluate the stationary state average value in the vicinity of the attraction, and the minimum value of $E(x)$ was denoted by $E_{a}$. Likewise, a section of $12~\text{m} \leq x \leq 18~\text{m}$ was selected for upstream of the attraction, and the minimum value of $E(x)$ in the section was indicated by $E_{up}$. For small values of $s$, the free flow phase was characterized by 
\begin{eqnarray}\label{eq:uni_free}
	E_{a} \approx 1\ \textrm{and}\ E_{up} \approx 1.
\end{eqnarray}
Likewise, the localized jam phase was characterized by 
\begin{eqnarray}\label{eq:uni_localized}
	0 < E_{a} < 1\ \textrm{and}\ E_{up} \approx 1.
\end{eqnarray}
When $Q$ and $s$ were large, the extended jam phase was characterized by 
\begin{eqnarray}\label{eq:uni_extended}
	0 \leq E_{a} < 1\ \textrm{and}\ 0 \leq E_{up} < 1. 
\end{eqnarray}
Figure~\ref{fig:results_3uni}(d) summarizes numerical results by dividing the parameter space of $Q$ and $s$ into different phases. Note that, in extended jam phase, one can observe the freezing phenomena with a certain probability.

\subsection{Bidirectional Flow}
\begin{figure}[!t]
	\centering
	\subfigure{}
	\includegraphics[width=10cm]{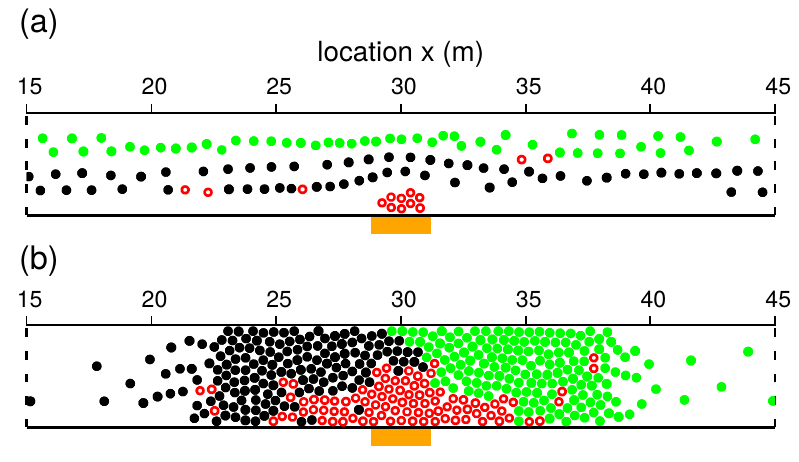}
	\centering
	\subfigure{}
	\includegraphics[width=5.5cm]{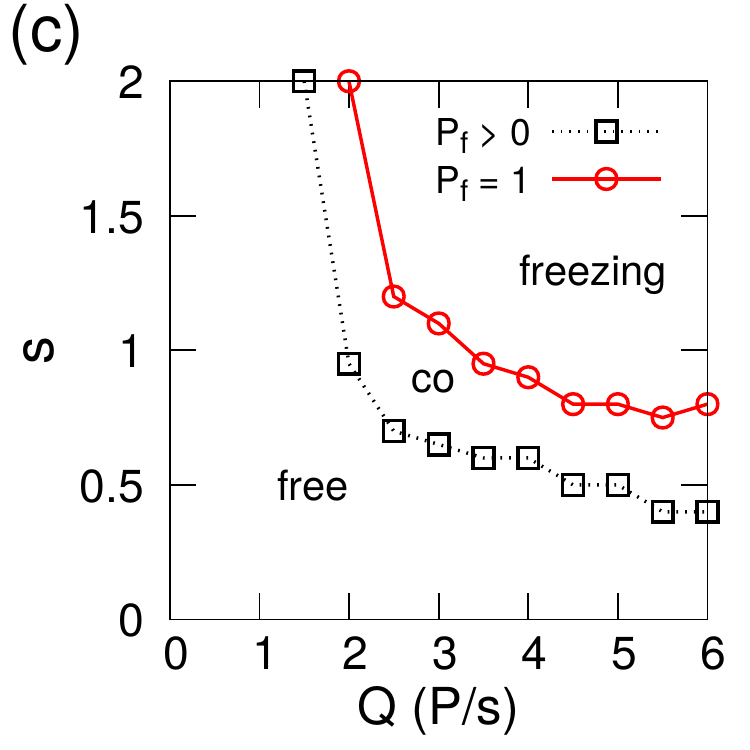}\vspace{-0.3cm}
	\caption{(color online) Representative snapshots of different passerby flow patterns in bidirectional flow: (a) free flow phase and (b) freezing phase. The attraction, depicted by an orange rectangle, is located at the center of the lower wall with open boundary conditions in the horizontal direction. Filled black and green circles indicate passersby walking to the right and to the left, respectively. Hollow red circles depict pedestrians attracted by the attraction. (c) Phase diagram summarizing the numerical results of bidirectional flow. Here, `free', `freezing', and `co' indicate the free flow phase, the freezing phase, and the coexisting phase, respectively.}
	\label{fig:results_3bi} 
\end{figure}

In bidirectional flow, the free flow phase appeared when $Q$ was small with weak $s$, similar to the case of the unidirectional flow scenario. From Figure~\ref{fig:results_3bi}(a), one can observe that passersby walk towards their destinations without being interrupted by the cluster of attracted pedestrians. Passersby walking to the right formed lines in the lower part of the corridor while the upper part of the corridor was occupied by passersby walking to the left. This spatial segregation appeared as a result of lane formation process which has been reported in previous studies~\cite{Feliciani_PRE2016, Helbing_PRE1995, Kretz_JSTAT2006_10a, Zhang_JSTAT2012}. At the same time, the attracted pedestrians formed a cluster in a semicircular shape near the attraction. If $Q$ and $s$ were large, a freezing phase was observed. In the freezing phase, oppositely walking pedestrians reached a complete stop because they blocked each other, as shown in Figure~\ref{fig:results_3bi}(b). 

One can identify the freezing phase by means of cumulative throughput at $x = 30$~m, according to Reference~\cite{Daganzo_1997}. The cumulative throughput at time $t$ is calculated by summing the number of passersby walking through $x = 30$~m from the beginning of the numerical simulation to time $t$. If the cumulative throughput does not change for $120$~s, it infers the appearance of the freezing phenomenon. The freezing probability $P_f$ was obtained by counting the occurrence of freezing phenomena over $50$ independent simulation runs for each parameter combination $(Q,\ s)$. For small value of $Q \leq 1.4$~P/s, $P_f$ is zero up to $s = 2$, indicating that the freezing phenomenon is not observable. Parameter combinations of $(Q, s)$ was categorized as the free flow phase if 
\begin{eqnarray}\label{eq:bi_free}
	P_f = 0,
\end{eqnarray}
yielding $E_{a} \approx 1$. The freezing phase was characterized by 
\begin{eqnarray}\label{eq:bi_freezing}
	P_f = 1
\end{eqnarray}
always showing $E_{a} = 0$. In addition, one can define a coexisting phase for parameter space between envelopes of $P_f = 0$ and $P_f = 1$. In the coexisting phase, either free flow or freezing phenomena can be observed depending on random seeds in the numerical simulations. That was characterized by  
\begin{eqnarray}\label{eq:bi_coexisting}
	0 < P_{f} < 1.
\end{eqnarray}
Figure~\ref{fig:results_3bi}(c) summarizes numerical results of bidirectional flow. In the coexisting phase, one can observe freezing phenomena with a certain probability.

\subsection{Microscopic Understanding of Jamming Transitions}
While previous subsections focused on describing collective patterns of various jam patterns, this subsection presents the appearance of such different patterns at the individual level in a unified way. The conflicts among pedestrians were closely looked into in order to understand the individual level behavior. Similar to previous studies~\cite{Kirchner_PRE2003, Nowak_PRE2012}, a conflict index was employed to measure the average number of conflicts per passerby. When two pedestrians are in contact and hinder each other, this situation is called a conflict. The number of conflicts $N_{c, i}(t)$ was evaluated by counting the number of pedestrians who hinder the progress of passerby $i$ at time $t$. In numerical simulations of Study~\Rmnum{3}, most conflicts appeared near the attraction; therefore, one can calculate the conflict index for pedestrians in location $x$ such that $25\text{~m} \leq x \leq 35\text{~m}$. The conflict index was measured as
\begin{eqnarray}\label{eq:ConflictIndex}
	n_c(t) = \frac{1}{|N_{p}(A, t)|} \sum_{i\in N_{p}(A, t)}{N_{c, i}(t)},
\end{eqnarray}
where $N_{p}(A, t)$ is the set of passersby in the section of $25\text{~m} \leq x \leq 35\text{~m}$ at time $t$ and $|N_{p}(A, t)|$ is the cardinality of the set $N_{p}(A, t)$.

\begin{figure}[!t]
	\centering
	\includegraphics[width=12cm]{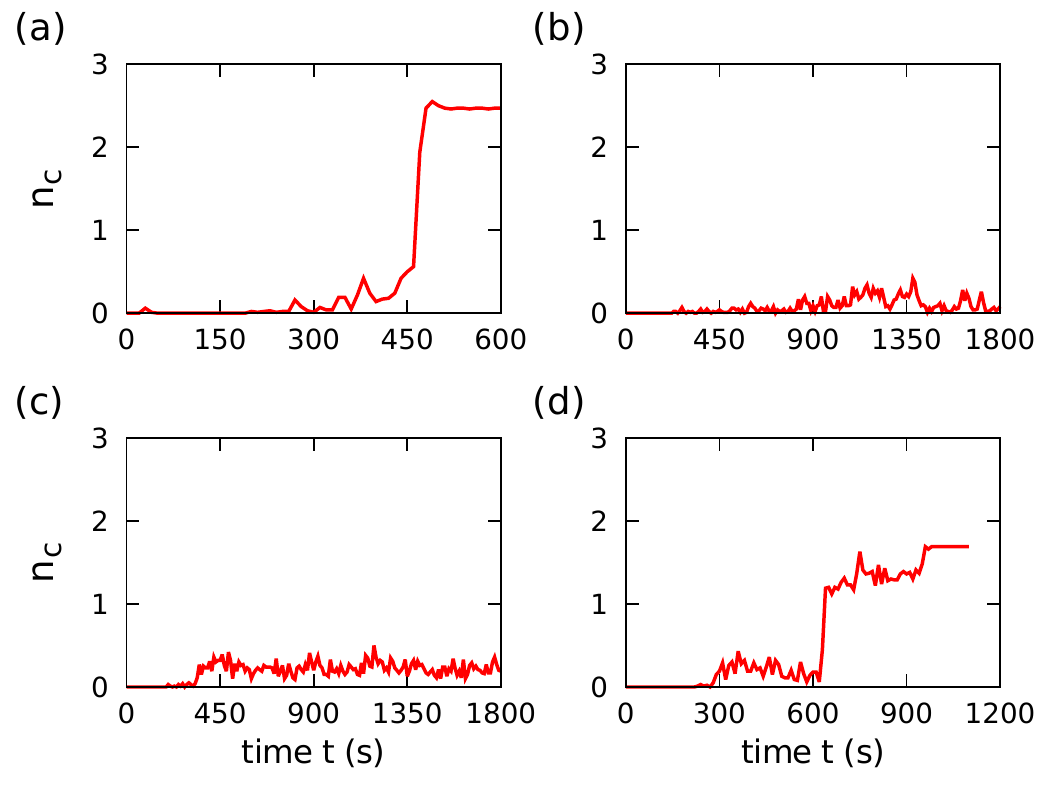}
	\caption{(Color online) Representative time series of conflict index $n_c(t)$ (a) freezing phase in bidirectional flow with $Q = 4$~P/s and $s = 1$. (b) localized jam phase in unidirectional flow with $Q = 5$~P/s and $s = 1$. (c) extended jam phase in unidirectional flow with $Q = 5$~P/s and $s = 1.8$. (d) same parameter combination $(Q, s)$ as (c), but with a different set of random seeds.}
	\label{fig:study3_conflict}
\end{figure}

The representative time series of conflict index $n_c(t)$ is presented in Figure~\ref{fig:study3_conflict}. As can be seen from Figure~\ref{fig:study3_conflict}(a), a sharp increase of the conflict index indicates the appearance of the freezing phenomenon, which leads the pedestrian flow into the freezing phase. In Figure~\ref{fig:study3_conflict}(b), the conflict index increases and then decreases in the course of time. One can observe the localized jam phase in which the jam near the attraction does not further grow upstream. Figures~\ref{fig:study3_conflict}(c) and \ref{fig:study3_conflict}(d) were generated with the same parameter combination $(Q, s) = (5, 1.8)$ in unidirectional flow but with different sets of random seeds. As shown in Figure~\ref{fig:study3_conflict}(c), in the extended jam phase, the conflict index $n_c(t)$ is maintained near a certain level after reaching the stationary state, indicating the persistent jam in the corridor. In Figure~\ref{fig:study3_conflict}(d), the behavior of $n_c(t)$ curve is similar to that of the extended jam phase in the beginning, but the curve abruptly increases at near $t = 600$~s. That is, the pedestrian flow eventually ends up in a freezing phenomenon in that conflicting pedestrians fail to coordinate their movements. 

In both bidirectional and unidirectional flows, attracted pedestrians often trigger conflicts among pedestrians. When the attracted pedestrians are walking towards the attraction, sometimes they cross the paths of passersby and hinder their walking. Furthermore, such crossing behavior of attracted pedestrians makes others change their walking directions due to the interpersonal repulsion, possibly giving rise to conflicts among the others. Once a couple of pedestrians hinder each other, they need some time and space to resolve the conflict by adjusting their walking directions. If there is not enough space for the pedestrian movement, the conflict situation cannot be resolved and it turns into a blockage in the pedestrian flow. Under higher pedestrian flux $Q$, conflicting pedestrians likely have less time for resolving the conflict while additional pedestrians arrive behind the blockage. Once the arriving pedestrians stand behind the blockage, the number of conflicts among pedestrians is rapidly increasing as indicated in Figures~\ref{fig:study3_conflict}(a) and \ref{fig:study3_conflict}(d). In the case of the bidirectional flow, this freezing phenomenon is similar to the freezing-by-heating phenomenon~\cite{Helbing_PRL2000}. However, note that the freezing phenomenon in the numerical simulations is caused by attracted pedestrians without noise terms in the equation of motion. 

\subsection{Attendee Cluster as a Dynamic Bottleneck}
As presented in previous subsections, the attendee cluster is acting as a pedestrian bottleneck for passersby. This finding is consistent with the description suggested by Goffman~\cite{Goffman_1971}. The flow through the bottleneck can show transitions from the free flow state to the jamming state and may end in gridlock. In the bidirectional flow, the free flow phase can turn into the freezing phase if $Q$ and $s$ are large. Jamming transitions in the unidirectional flow are different from those of the bidirectional flow: from the free flow phase to the localized jam phase, and then to the extended jam phase. In addition, it is possible that the extended jam phase ends up in freezing phenomena for large $Q$ and $s$. Although different jamming transitions were observed for uni- and bi-directional flow scenarios, an attendee cluster can be conceptualized as a dynamic bottleneck in the sense that its size changes over time according to the joining behavior of attracted pedestrians. One might assume that an attendee cluster can be approximated as a static bottleneck which is at a fixed location and whose size does not change over time. Readers can find representative studies on static bottlenecks from Hoogendoorn and Daamen~\cite{Hoogendoorn_TrSci2005} and Seyfried~\textit{et al.}~\cite{Seyfried_TrSci2009}. However, the assumption of the static bottleneck cannot reflect the fluctuating attendee cluster size, which arises from the interactions among the attraction, attracted pedestrians, and passersby. 

\begin{figure}[!t]
	\centering
	\begin{tabular}{ccc}
		\includegraphics[width=4.5cm]{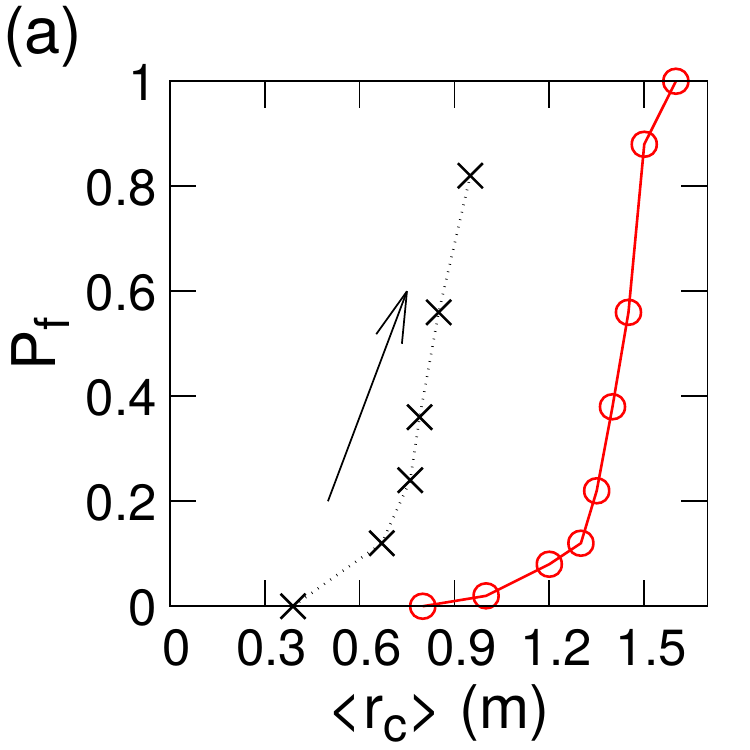}&\hspace{-0.5cm}
		\includegraphics[width=4.5cm]{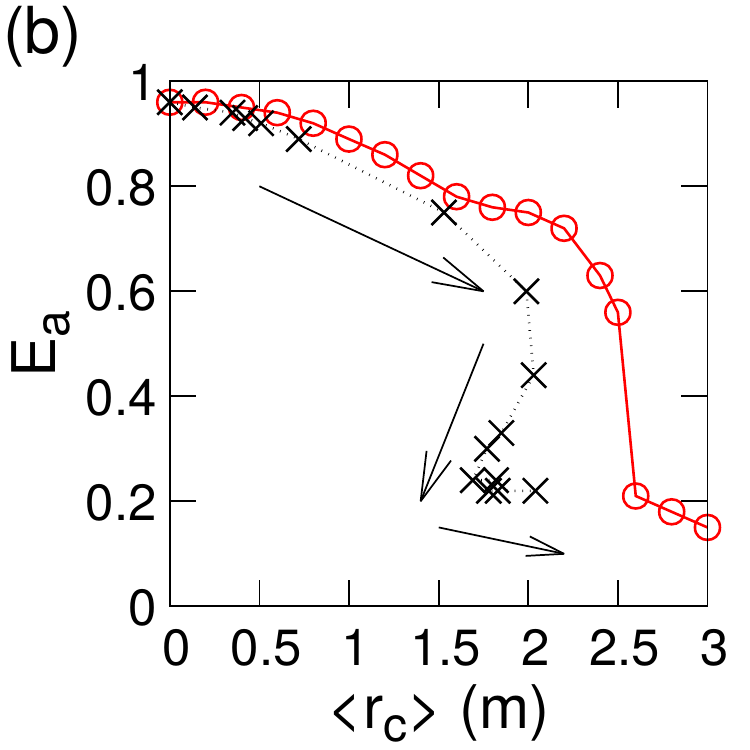}&\hspace{-0.5cm}
		\includegraphics[width=4.5cm]{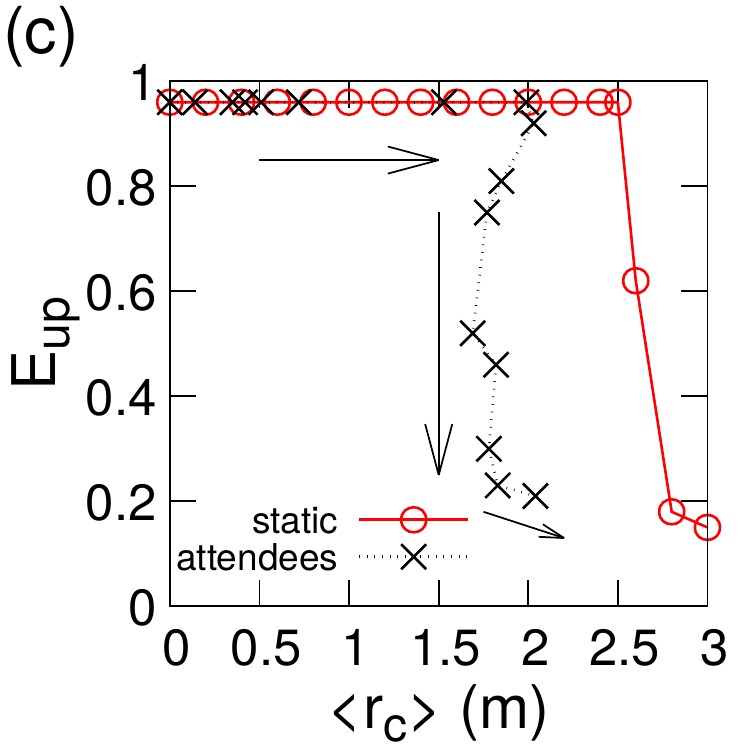}\vspace{-0.2cm}
	\end{tabular}
	\caption{(Color online) Dependence of various measures on a stationary state average of cluster size $\langle r_c \rangle$ for pedestrian influx $Q = 5$~P/s. The results of a static bottleneck and an attendee cluster are denoted by $\circ$ and $\times$, respectively. Arrows indicate the direction of increasing $s$, for the results of the attendee cluster. (a) Freezing probability $P_f$ for bidirectional flow indicating the appearance of freezing phase (b) local efficiency near the attraction $E_{a}$ for unidirectional flow reflecting the onset of localized jam phase, and (c) local efficiency upstream $E_{up}$ for unidirectional flow, which is relevant to extended jam phase.}
	\label{fig:study3_bottleneck} 
\end{figure}

In order to understand the influence of the attendee cluster size on jamming transitions, numerical simulations were performed with a static bottleneck. In this case, all the pedestrians were modeled as passersby. In doing so, one can exclude the interactions among passersby and attendees, thereby focusing on the influence of reduced available space. For the comparison, a stationary state average of cluster size $\langle r_c \rangle$ was used because the attendee cluster size changes in the course of time but the size of the static bottleneck is constant. In the case of a static bottleneck, the notation of $\langle r_c \rangle$ was also used for convenience. A semicircle with radius $\langle r_c \rangle$ was placed at the center of the lower corridor boundary, acting as a static bottleneck. By changing $\langle r_c \rangle$, one can observe the behavior of various measures including $P_f$, $E_{a}$, and $E_{up}$ (see Figure~\ref{fig:study3_bottleneck}). 

It was obvious that larger $\langle r_c \rangle$ led to higher freezing probability $P_f$ for the bidirectional flow, as shown in Figure~\ref{fig:study3_bottleneck}(a). However, $P_f$ of the attendee cluster case was higher than that of the static bottleneck for a given value of $\langle r_c \rangle$. While $E_{a}$ and $E_{up}$ curves obtained from the static bottleneck case showed a clear dependence on $\langle r_c \rangle$, those from attendee cluster did not show clear tendency when $\langle r_c \rangle > 1.5$~m (see Figures~\ref{fig:study3_bottleneck}(b) and \ref{fig:study3_bottleneck}(c)). Although increasing $\langle r_c \rangle$ evidently led to a localized jam transition, it can be suggested that conflicts among pedestrians play an important role in jamming transitions if $\langle r_c \rangle$ is large enough. It can, therefore, be said that the attendee cluster acts as a dynamic bottleneck that behaves qualitatively different than the static bottleneck in terms of jamming transitions. 

\chapter{Conclusions and Discussions}
\label{sec:Conclusions}
\section{Summary}
This dissertation aimed at studying collective dynamics of pedestrians interacting with attractions. The dissertation topic was investigated based on three research questions presented in Section~\ref{sec:RQ}:
\begin{itemize}[label={}]
	\item \textbf{(RQ1)} how do collective patterns of pedestrian motions emerge from the attractive interactions between pedestrians and attractions?
	\item \textbf{(RQ2)} how can social influence on one's choice behavior shape the collective patterns of pedestrians' visiting behavior?
	\item \textbf{(RQ3)} how can pedestrian flow interacting with an attraction result in pedestrian jams?
\end{itemize}
In other words, \textbf{RQ1} mainly focused on the formation of attendees near attractions, and \textbf{RQ2} paid particular attention to collective visiting patterns when individual choice behavior is likely influenced by others. \textbf{RQ3} devoted to pedestrian jamming transitions induced by an attraction. 

In order to address the research questions, microscopic pedestrian models were extended by incorporating the attractive interactions between pedestrians and attractions, and implemented in numerical simulations. After performing the numerical simulations, various collective patterns were identified and summarized in phase diagrams based on macroscopic measures. By doing so, this dissertation investigated the dynamics of pedestrian flow interacting with attractions from the perspective of attracted pedestrians and passersby. 

In line with \textbf{RQ1}, Study~\Rmnum{1} addressed collective effects of attractive interactions between pedestrians and attractions by means of numerical simulations. In order to do that, the attractive interactions were modeled [see Equation~(\ref{eq:AR})] and appended to the social force model [see Equation~(\ref{eq:SFM_study1})]. By means of numerical simulations, the attractive interactions were examined by controlling the attractive interaction strength and the pedestrian density. The interactions led pedestrians to form stable clusters around attractions, or even to rush into attractions if the interaction becomes stronger. In this case, the attracted pedestrians tended to push each other because of the interpersonal repulsion effect. It was also found that for high pedestrian density and intermediate interaction strength, some pedestrians rush into attractions while others move to neighboring attractions. The moving pedestrians were also attracted by the attractions but cannot stay around the attraction. This is due to the interpersonal repulsion effect by other pedestrians closer to the attractions. These collective patterns of pedestrian movements or phases and transitions between them were systematically presented in a phase diagram, see Figure~\ref{fig:results_1}(e). 

With reference to \textbf{RQ2}, Study~\Rmnum{2} investigated the social influence on collective visiting behavior by developing a joining probability model as shown in Equation~(\ref{eq:P_a}). The joining probability was formulated as a function of social influence from others, reflecting that individual choice behavior is likely influenced by others. Numerical simulations produced different patterns of pedestrian behavior depending on the strength of the social influence and the average duration of visiting an attraction. When the social influence was strong along with a long duration of visiting an attraction, the saturated phase was defined at which all the pedestrians have visited the attraction. If the social influence was not strong enough, the unsaturated phase appeared where one can observe that some pedestrians head for the attraction while others walk in their desired direction. These collective patterns of pedestrian behavior were summarized in a phase diagram [see Figure~\ref{fig:results_2}(c)] by comparing the number of pedestrians who have visited the attraction to the number of pedestrians near the attraction. One can identify under what conditions enhancing the social influence strength and the average duration of visiting an attraction would be effective by measuring the marginal benefits with respect to those parameters.

Motivated by \textbf{RQ3}, Study~\Rmnum{3} numerically studied jamming transitions in pedestrian flow interacting with an attraction, mostly based on the social force model for pedestrians who can join the attraction. Various pedestrian flow patterns were observed by controlling pedestrian influx and the social influence parameter. For bidirectional flow scenario, one could observe a transition from the free flow phase to the freezing phase in which oppositely walking pedestrians reach a complete stop and block each other. On the other hand, a different transition behavior appeared in the unidirectional flow scenario, i.e., from the free flow phase to the localized jam phase, and then to the extended jam phase. It was also observed that the extended jam phase could end up in freezing phenomena with a certain probability when pedestrian flux was high with strong social influence. Study~\Rmnum{3} highlighted that attractive interactions between pedestrians and an attraction could lead to jamming transitions due to the conflicts among pedestrians near the attraction. 

\section{Contributions}
This dissertation contributes to the current body of knowledge on the collective behavior of pedestrian motions. This is achieved by modeling their dynamics interacting with attractions and providing possible explanations of the collective patterns. The contributions of this dissertation are further stated in three folds.

In Study~\Rmnum{1}, the attractive force model was developed based on the idea of ``short-range strong repulsive and long-range weak attractive interactions \cite{DOrsogna_PRL2006, Mogilner_MathBio2003}.'' While previous studies~\cite{DOrsogna_PRL2006, Mogilner_MathBio2003} developed attractive interaction models for interacting self-propelled particles, Study~\Rmnum{1} presented an attractive interaction model for pedestrians interacting with attractions. The presented model demonstrated that the social force models can be extended to predict various collective behaviors associated with attractions. The appearance of various collective patterns was explained by the interplay between attraction strength and interpersonal repulsion effect. Interestingly, despite its simple formulation, the presented model can provide a plausible explanation of extreme pedestrian behavior in stores, such as Black Friday incidents in the United States. 

In Study~\Rmnum{2}, the joining probability model was developed in line with selective attention~\cite{Goldstein_2007, Wickens_1999} and social influence~\cite{Bearden_JCR1989, Childers_JCR1992, Gallup_PNAS2012, Kaltcheva_JMkt2006, Milgram_JPSP1969}. Previous studies ~\cite{Gallup_PNAS2012, Milgram_JPSP1969} reported that the joining probability increases as the stimulus group size grows. The presented joining probability model in Study~\Rmnum{2} was developed by incorporating with social influence parameter and duration of visiting an attraction that were not considered in the previous studies. The numerical simulation results showed different patterns of collective visiting behavior: unsaturated and saturated phases. The appearance of the saturated phase was explained by an interplay between social influence strength and the average duration of visiting an attraction.

Study~\Rmnum{3} characterized the dynamics of pedestrian jams induced by the attraction. By describing pedestrian conflicts at a microscopic level, it was highlighted that attractive interactions between pedestrians and an attraction could trigger jamming transitions by increasing the number of conflicts among pedestrians near the attraction. Therefore, it is suggested that existence of an attraction acts similar to a noise term in an equation of particle motions~\cite{Helbing_PRL2000}, which leads to freezing phenomena in the pedestrian flow. Furthermore, an attendee cluster near an attraction was conceptualized as a dynamic bottleneck and identified that a dynamic bottleneck behaves qualitatively different than the static bottleneck in terms of jamming transitions.

The findings presented in this dissertation can provide an insight into pedestrian flow patterns in stores and pedestrian facility management strategies. Numerical simulation results imply that safe and efficient use of pedestrian facilities can be achieved by moderating the control variables. For instance, findings from Study~\Rmnum{2} suggest that increasing social influence strength $s$ and the average duration of visiting an attraction $t_d$ is not effective all the time, especially when these variables are already high enough. Increasing relative attraction strength $C$ and pedestrian density $\rho$ can even lead to extreme and dangerous situations such as the competitive phase in Study~\Rmnum{1}. Likewise, increasing pedestrian influx $Q$ and social influence strength $s$ yields undesirable situations, like freezing phenomena in Study~\Rmnum{3}, in terms of pedestrian safety and efficiency.

\section{Limitations}
Although the presented models were developed based on analogies with self-propelled particles~\cite{Mogilner_MathBio2003, Vicsek_PRL1995} and findings in behavioral science studies~\cite{Gallup_PNAS2012, Milgram_JPSP1969}, numerical simulation results were not compared against field observations yet. For the place where impulse stops are significant, the presented model in Study~\Rmnum{1} can be tested and parameters in the attractive force model (i.e., $C_a$, $C_r$, $l_a$, and $l_r$) need to be measured from field observations. For the joining behavior model in Studies~\Rmnum{2} and \Rmnum{3}, the average duration of visiting an attraction $t_d$ can be measured for attendees near an attraction and then social influence parameter $s$ can be predicted based on the number of pedestrians having visited the attraction. The predicted value of $s$ can be utilized to test various scenarios. 

In this dissertation, simple scenarios of pedestrian flow in a straight corridor have been focused in order to study essential features of collective pedestrian dynamics interacting with attractions. Although a straight corridor can be assumed as a basic element of pedestrian facilities, some pedestrian flow scenarios such as crossings at intersections and multi-directional flows in open space cannot be covered by the presented results of the numerical simulations. Therefore, suitable modifications and extensions are required to apply the presented models to real world examples. 

For simplicity, it was assumed that the pedestrians in numerical simulations did not change their decision after they decided what they were going to do. However, pedestrians in the real world can change their decision as a response to the situations around them. A pedestrian, who decides to join an attraction, can change one's mind after realizing that there are too many attendees near the attraction. Modeling and implementing such decision changing behavior can be an important extension of the presented models. 

\section{Future Work}
The presented models can be further improved and extended. The attractive force model in Study~\Rmnum{1} was developed in line with the social force model and predicted various collective patterns arising from the attractive interactions between pedestrians and attractions. However, the presented model appears to be valid only for the social force model. Although the social force model and other force-based models have well explained various collective patterns, it has been pointed out that the force-based models are complicated and a careful implementation is required in order to avoid the pitfalls~\cite{Chraibi_PRE2010, Koster_PRE2013, Kretz_PhysicaA2015}. Chraibi~\textit{et al.}~\cite{Chraibi_PRE2015} remarked that velocity-based models are gaining more attention among researchers because of their simplicity. Developing attractive interaction model for velocity-based models can be a topic for future work. 

The collective phenomena predicted in Study~\Rmnum{1} can be compared against field observations. In Study~\Rmnum{1}, it was assumed that all the pedestrians were subject to the same attractive interactions. A future study might explore trajectory data of pedestrians fighting over merchandise during shopping holidays like Black Friday. In that case, pedestrians are likely to behave in the same way near an attraction, similar to the setup of numerical simulations in Study~\Rmnum{1}. However, one can easily observe that an attraction is not always attractive for everyone. For instance, in a museum, some visitors are watching a portrait because they are interested in it while others pass by it and see other artworks. Consequently, verification of Study~\Rmnum{1} can be performed for pedestrians attracted by attractions, inferring that those pedestrians are subject to the same attractive interaction. Attracted pedestrian behavior near artworks in museums and shop displays in shopping malls can be studied based on observed trajectories and velocity field. Note that the presented model in Study~\Rmnum{1} predicts pedestrian behavior under the influence of attractions for a short period, mainly focusing on momentary response to the attractions.  

The joining behavior model in Study~\Rmnum{2} can be formulated in different forms. The presented joining behavior model was developed based on the idea of behavioral contagion, inferring that an individual tends to follow what others do~\cite{Gallup_PNAS2012, Milgram_JPSP1969}. Although the assumption of behavioral contagion has been widely applied especially in pedestrian evacuation studies~\cite{Helbing_Nature2000}, the assumption has been questioned. Haghani and Sarvi~\cite{Haghani_AnimalBehaviour2017} reported that evacuees showed congestion avoidance behavior in emergency evacuations especially when evacuees were well aware of the situations near exits. That is, the behavioral contagion is not always useful for explaining pedestrian joining behavior. If there are too many attendees near an attraction, the attraction is not necessarily attractive for every pedestrian near the attraction. For some pedestrians, seeing too many attendees can decrease the probability of joining the attraction. In future studies, the congestion avoidance behavior can be incorporated in the joining behavior model. In addition, the joining behavior model can be extended in order to take into account various characteristics of pedestrians and attractions by adding additional behavioral features, such as the explicit representation of group behaviors~\cite{Zanlungo_PRE2014} and interest function~\cite{Kielar_SMPT2016}. One can postulate that pedestrians have time budget, so they evaluate the attractiveness and the cost of time due to joining the attractions. Heterogeneous properties of pedestrian joining behavior can be considered in terms of the congestion avoidance behavior, group behaviors, interest functions, and the cost of time. 

In future studies, the presented joining behavior model can be verified against empirical observations. It was assumed that the joining probability grew as the number of attendees increased and the pedestrians decided to join the attraction 10 m ahead of it. In order to obtain naturalistic movement of pedestrians near attractions, it is desirable to obtain video recordings of public areas like shopping malls. Pedestrian trajectories need to be extracted from the video recordings. The extracted trajectories can be used for estimating the distance between pedestrians and attractions, and for identifying decision moments at which pedestrians suddenly change their walking direction. Based on the decision moments, one can understand when and where pedestrians decide to join the attraction. From the video recordings, the number of attendees near attractions and passersby can be counted and the joining probability can be measured for different attendee size at various places and time. One can discover the relationship between the joining probability and the number of attendees near the attractions. If the joining probability increases as the number of attendees grows, one can say that the assumption of behavioral contagion is observable. On the other hand, if a decreasing trend of joining probability against the number of attendees is observed, the congestion avoidance behavior might exist. In addition, the duration of visiting an attraction was assumed to follow an exponential distribution. The duration of visiting each attraction can be measured for each attendee, so its distribution can be identified. The evacuation exit choice models~\cite{Haghani_AnimalBehaviour2017, Wagoum_RSOS2017} can be extended and modified for verifying assumptions of the presented joining probability model.

A natural progression of Study~\Rmnum{3} is to analyze the numerical simulation results from the perspective of capacity estimation. Capacity estimation can focus on the optimal capacity, balancing the mobility needs for passersby and the activity needs for attracted pedestrians. One can explicitly consider the capacity of the attractions, meaning that only a certain number of attendees can stay near the attractions. The concept of stochastic capacity~\cite{Geistefeldt_ISTTT2009, Kerner_PRE2014_vol89, Minderhoud_TRR1997} can also be studied based on the findings of Study~\Rmnum{3}. In Study~\Rmnum{3}, for some parameter values, speed breakdown is observed depending on random seeds, inferring that capacity might follow a probability distribution.

Future extensions of Study~\Rmnum{3} can be planned from the perspective of pedestrian flow experiments. Although the joining behavior presented in this study might not be controlled in experimental studies, the experiments can be performed for different levels of pedestrian flux and joining probability. The number of passersby and pedestrians joining the attraction, and their trajectories with velocity vectors need to be collected from future pedestrian experiments. For various experiment configurations, the number of conflicts among pedestrians can be measured and the influence of the conflicts on pedestrian jams can be analyzed.

In this dissertation, the parameter values used in numerical simulations were in a limited range. Although the selected parameter values produced interesting collective phenomena, performing additional numerical simulations with a larger set of their values might yield more findings that are interesting. It is noted that the shape of phase diagrams presented in the dissertation can be changed for different parameter values and numerical simulation settings such as corridor width and the number of attractions. Changing numerical simulation parameters might show the appearance of different collective patterns of pedestrian motions. 

\renewcommand{\bibname}{References}
\bibliographystyle{plain} 


\addpublication{Kwak, J., Jo, H-H., Luttinen, T., and Kosonen, I}{Collective dynamics of pedestrians interacting with attractions}{Physical Review E}{Volume 88, 6, 062810}{December}{2013}{American Physical Society}{j1}
\addcontribution{Jaeyoung Kwak is the first/corresponding author of Publication~\Rmnum{1}. He was responsible for conducting all the research work including development of research ideas/questions, formulation and implementation of the presented model, and analysis of the numerical simulation results. The manuscript was mainly written by Kwak. Dr. Jo actively advised Kwak on writing the manuscript including the manuscript structure and style, and preparing figures and tables. Prof. Luttinen and Dr. Kosonen were in a commentary role.}
\addpublicationpdf{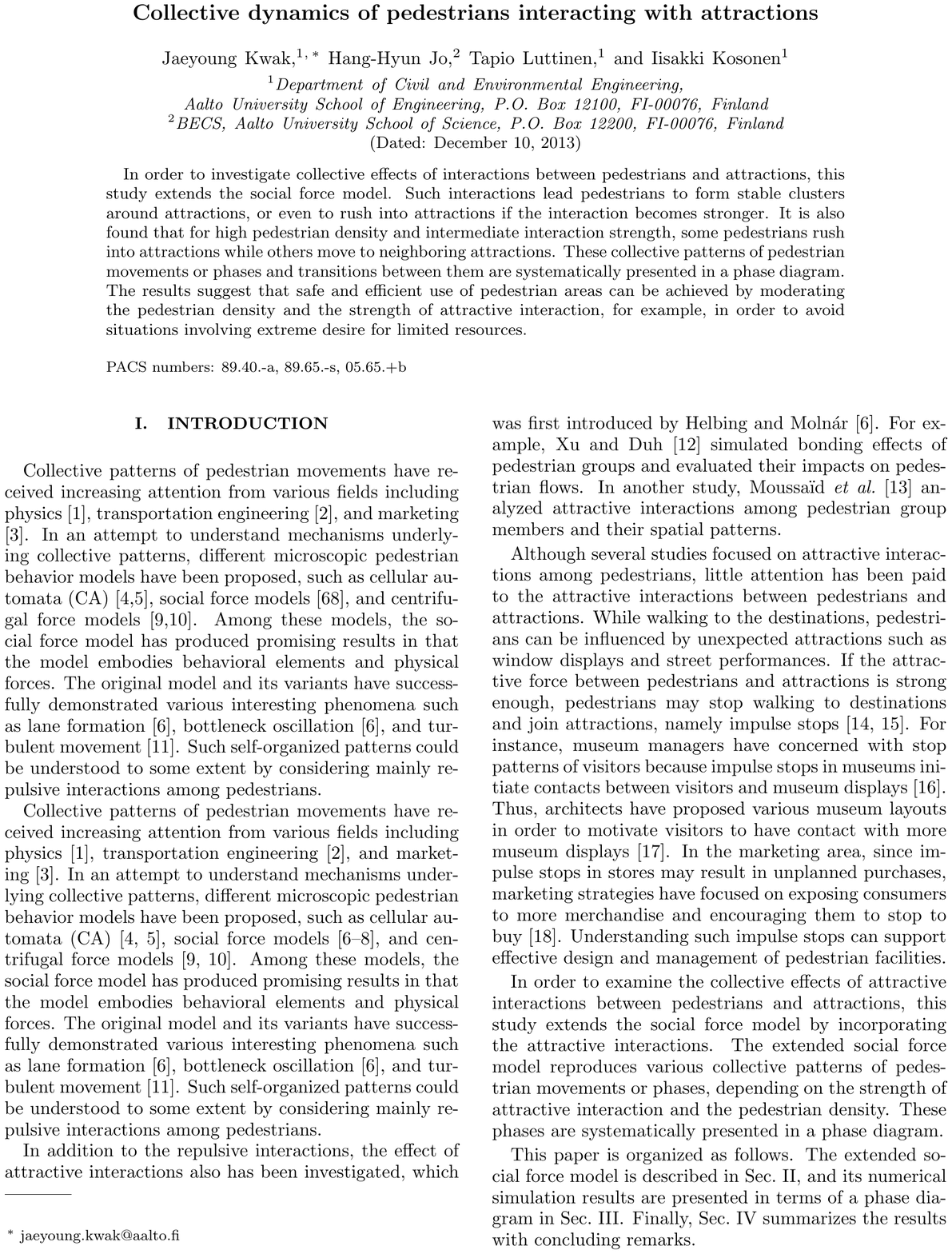}
%
\addpublication{Kwak, J., Jo, H-H., Luttinen, T., and Kosonen, I}{Effects of switching behavior for the attraction on pedestrian dynamics}{PLOS ONE}{Volume 10, 7, e0133668}{July}{2015}{Authors}{j2}
\addcontribution{Jaeyoung Kwak is the first/corresponding author of Publication~\Rmnum{2}. He was responsible for conducting all the research work. Kwak conceived the research ideas/questions, designed and performed the numerical simulations, and analyzed the simulation results. Dr. Jo actively advised Kwak on developing the research ideas/questions, designing the numerical experiments, and analyzing the results. The manuscript was mainly written by Kwak, and other authors (Dr. Jo, Prof. Luttinen, and Dr. Kosonen) provided feedback for the manuscript.}
\addpublicationpdf{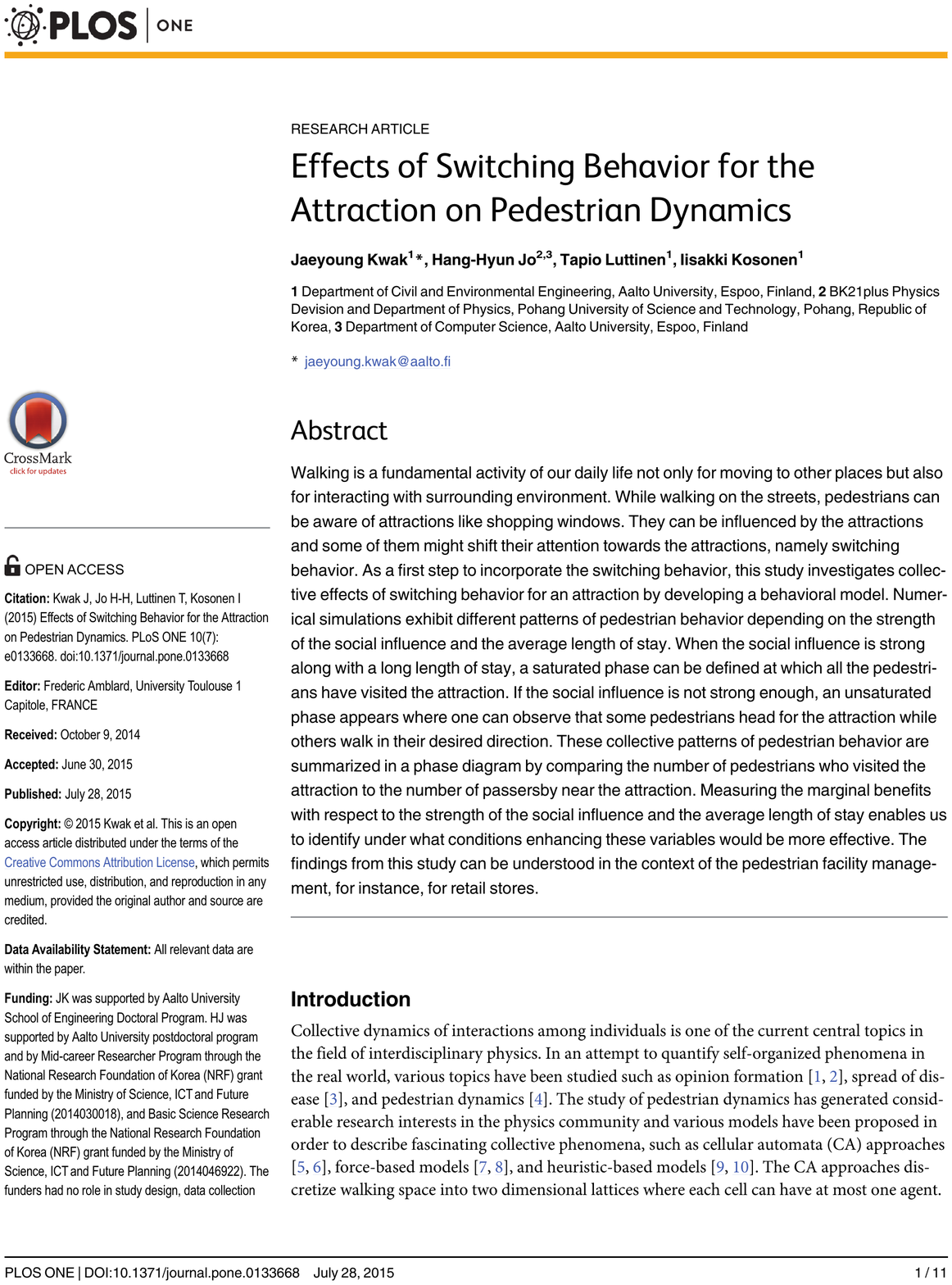}
%
\addpublication{Kwak, J., Jo, H-H., Luttinen, T., and Kosonen, I}{Jamming transitions induced by an attraction in pedestrian flow}{Physical Review E}{Volume 96, 2, 022319}{August}{2017}{American Physical Society}{j3}
\addcontribution{Jaeyoung Kwak is the first/corresponding author of Publication~\Rmnum{3}. He was responsible for conducting all the research work. He conceived the research ideas and developed the research questions. Kwak formulated the presented model, performed the numerical simulations, and interpreted the results. The manuscript was mainly written by Kwak, and other authors (Dr. Jo, Prof. Luttinen, and Dr. Kosonen) were in a commentary role.}
\addpublicationpdf{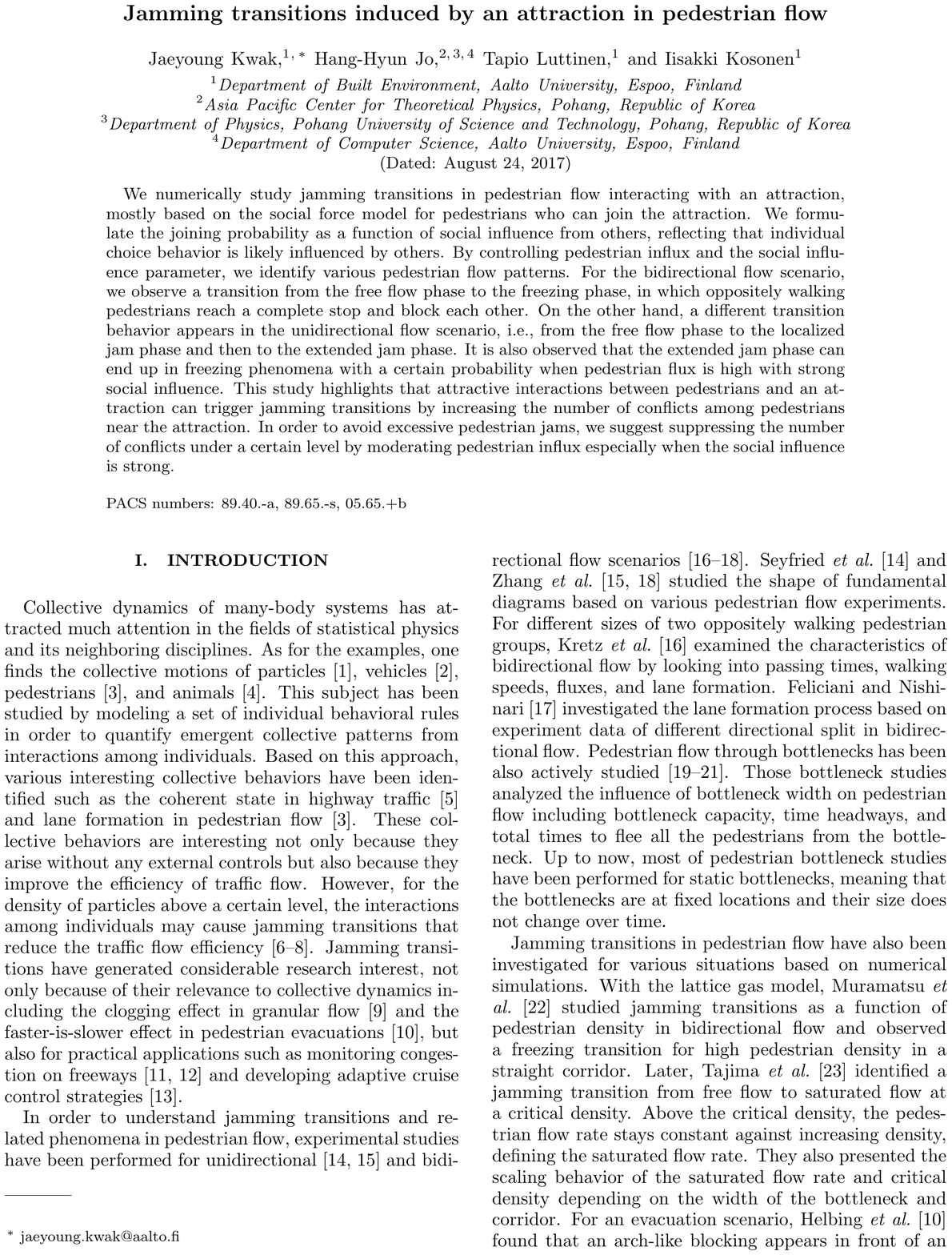}

\end{document}